\documentclass[%
reprint, 
superscriptaddress,
amsmath,amssymb,
aps,
prx,
]{revtex4-2}

\usepackage{graphicx}
\usepackage{picture}
\usepackage{placeins}
\usepackage{float}
\DeclareGraphicsExtensions{.pdf,.eps} 
\usepackage{amsmath,scalefnt}
\usepackage{mathtools}
\usepackage{amssymb}
\usepackage{verbatim}
\usepackage{amsmath,amsfonts,bbm}
\usepackage{amsbsy}
\usepackage{color}
\usepackage{cancel}
\usepackage{soul}
\usepackage{dsfont}
\usepackage[svgnames]{xcolor} 
\usepackage[compat=1.0.0]{tikz-feynman}
\usepackage[english]{babel}

\usepackage{hyperref}
\hypersetup{breaklinks=true,colorlinks=true,linkcolor=DarkBlue,urlcolor=DarkBlue,citecolor=DarkBlue}

\usepackage{subfigure}

\usepackage{braket}
\newcommand{\da}[0]{^{\dagger}}

\def\bs#1{\boldsymbol{#1}}
\def\ba#1{\left(\begin{array}{#1}}
	\def\ea{\end{array}\right)}
\def\bsm{\left(\begin{smallmatrix}}
	\def\esm{\end{smallmatrix}\right)}
\def\unit#1{\, \mathrm{#1}}

\def\bs#1{\boldsymbol{#1}}

\begin{document}


\title{Non-Gaussianity as a  signature of a quantum theory of gravity}


\author{Richard Howl}
\email{rjhowl@gmail.com}
\affiliation{School of Mathematical Sciences,
	University of Nottingham,
	University Park,
	Nottingham, NG7 2RD,
	United Kingdom}\thanks{R.H.\ was affiliated with$^{\,1}$ when the article was submitted for publication.}
\affiliation{QICI Quantum Information and Computation Initiative, Department of Computer Science, The University of Hong Kong, Pokfulam Road,
	Hong Kong}\thanks{R.H.\ was affiliated with$^{\,2,\,3}$ when the article was accepted for publication.}
\affiliation{Quantum Group, Department of Computer Science, University of Oxford, Wolfson Building, Parks Road, Oxford, OX1 3QD, United Kingdom}\thanks{R.H.\ was affiliated with$^{\,2,\,3}$ when the article was accepted for publication.}

\author{Vlatko Vedral}
\affiliation{Clarendon Laboratory,
	Department of Physics, 
	University of Oxford,
	Oxford, OX1 3PU,
	United Kingdom}
\affiliation{Centre for Quantum Technologies,
	National University of Singapore,
	Block S15,
	3 Science Drive 2, Singapore 117543}

\author{Devang Naik}
\affiliation{LP2N, Laboratoire Photonique, Num\'{e}rique et Nanosciences, Universit\'{e} Bordeaux-IOGS-CNRS:UMR 5298, F-33400 Talence, France}

\author{Marios Christodoulou}
\affiliation{QICI Quantum Information and Computation Initiative, Department of Computer Science, The University of Hong Kong, Pokfulam Road,
	Hong Kong}
\affiliation{Clarendon Laboratory,
	Department of Physics, 
	University of Oxford,
	Oxford, OX1 3PU,
	United Kingdom}

\author{Carlo Rovelli}
\affiliation{CPT,
	Aix-Marseille Univ, 
	Universit\'{e} de Toulon,
	CNRS,
	F-13288 Marseille,
	France}
\affiliation{Perimeter Institute, 31 Caroline Street N, Waterloo ON, N2L 2Y5, Canada}
\affiliation{The Rotman Institute of Philosophy, 1151 Richmond St.\ N London,  N6A 5B7, Canada}

\author{Aditya Iyer}
\affiliation{Clarendon Laboratory,
	Department of Physics, 
	University of Oxford,
	Oxford, OX1 3PU,
	United Kingdom}

\begin{abstract} 
	Table-top tests of quantum gravity (QG) have long been thought to be practically impossible. However,  remarkably, due to   rapid progress in quantum information science (QIS), such tests may soon be achievable. Here, we uncover an exciting new theoretical link between QG and QIS that also leads to a radical new way of testing QG with QIS experiments. Specifically, we find that only a quantum, not classical, theory of gravity can create non-Gaussianity, a QIS resource that is necessary  for universal quantum computation, in the quantum field state of matter. This allows for  tests based on QIS in which non-Gaussianity in matter is used as a signature of QG.	In comparison to previous studies of testing QG with QIS where entanglement is used to witness  QG when all other quantum interactions are excluded,  our non-Gaussianity witness cannot be created by direct classical gravity interactions, facilitating  tests  that are not constrained by the existence of such processes.  Our new signature of QG also enables tests that are based on just a single  rather than  multi-partite quantum system,  simplifying previously considered experimental setups.  We describe a   table-top test of QG that uses our non-Gaussianity signature and which is based on just a single quantum system, a Bose-Einstein condensate (BEC), in  a single location. In contrast to proposals based on opto-mechanical setups, BECs have already been manipulated into massive non-classical states, aiding the prospect of testing QG in the near future. 
\end{abstract}
 
 
 
\maketitle

\section{Introduction} \label{sec:Intro}

Shortly after Einstein formulated general relativity (GR) he wondered how quantum theory (QT) would modify it \cite{EinsteinQG}. Yet, over a hundred years later, there is still no consensus on how these two fundamental  theories should be unified \cite{isham1995structural,rovelli1998strings,rovelli2000notes,oriti2009approaches,QGABriefHistory}. The conventional approach is to apply the principles of QT to gravity \cite{penrose2014gravitization}, resulting in a quantum gravity (QG) theory, such as string theory \cite{superstring1,superstring2,Polchinski1,Polchinski2} or loop QG \cite{rovelli1988new,rovelli2004quantum,thiemann2008modern}. However, since it is not as straightforward to apply QT  to gravity as compared with the other fundamental forces \cite{ROSENFELD1963353,Carlip_2008}, an alternative class of unifying theories has been developed,  classical gravity (CG) theories, such as semi-classical gravity \cite{ROSENFELD1963353,moller1962theories,kibble1978relativistic,Kibble_1980}, where matter is quantized but gravity remains fundamentally classical \cite{Carlip_2008}.

The hope has been that theoretical study alone would lead us to how GR and QT are unified in nature. However, the fact that there are several proposals illustrates that this is unlikely to happen and that  experimental intervention is required \cite{Carlip_2008}. Until  recently, the common view was that there is little hope of laboratory tests of QG since we  need to probe GR near a small length scale, the Planck length, where QT effects of spacetime become relevant \cite{Note1}, but for which we would likely need to build a Milky-Way-sized particle accelerator\footnotetext[1]{Rather than using laboratory settings, such as a particle accelerator, it may also be possible to probe general Planck length/energy scale effects of  QG through cosmological and astrophysical studies \cite{BARRAU2017189} - see Section \ref{sec:CMB} for a brief overview. However, these  lack the control of laboratory settings and, because of this, the predicted QG effects tend  to also be explainable  by classical theories of gravity \cite{QDiscordCMB}. There are also proposals for tests of specific phenomenological models of QG, rather than fundamental theories of QG, that are designed to show possible Planck length/energy scale effects, such as  violations of Lorentz-invariance, at accessible levels in particular experiments \cite{PhenomQG}.} \cite{hossenfelder2018lost}. However,  there is another important scale, the Planck mass scale, where gravitational effects of massive quantum systems become relevant, allowing us, in particular, to distinguish QG from CG \cite{RovelliQGExp}. This mass scale should be within reach soon in laboratory settings due to the rapidly developing field of  quantum information science (QIS) \cite{Ulbricht20015,ReviewPaper}. This has led to several proposals being  developed recently for tests of QG using  techniques of QIS, see e.g.\ \cite{kafri2013noise,Ulbricht20015,Carlesso_2019,BoseQGExp,BoseLocal,VedralQGExp,PhysRevD.98.046001,PhysRevLett.119.120402,Kafri_2014,krisnanda2020observable}. Of these, a particularly  promising   experimental proposal is the Bose-Marletto-Vedral (BMV) experiment \cite{BoseQGExp,BoseLocal,VedralQGExp,PhysRevD.98.046001} where, under the condition that   all other quantum interactions can be excluded, the  creation of entanglement between two microspheres, each in a superposition of two locations, is used as a witness of QG. Due to the strength of this effect, and the hope of mesoscopic  superposition states in opto-mechanical systems \cite{PhysRevLett.117.143003,SuperconductingMicrosphere}, it is thought that this QIS-inspired experiment of QG could be possible  in the near future \cite{BoseQGExp,BoseLocal,VedralQGExp,PhysRevD.98.046001}.


An issue with an entanglement-based test of QG, however, is that classical, as well as quantum theories of gravity can create entanglement. For example, modes of a quantum field can become entangled by a classically expanding universe \cite{PhysRevLett.21.562,PhysRev.183.1057,PhysRevD.3.346,BirrelandDavies,Entanglementinthesecondquantizationformalism,BALL2006550,PhysRevD.82.045030,PhysRevD.89.024022}. This has resulted in questions concerning the reliability of using entanglement  as a witness of QG \cite{Hall_2018,pal2019experimental,Reginatto_2019,marletto2019answers}. In particular, in the BMV proposal, entanglement as a witness of QG is based on the assumption that CG acts as a local operations and communication channel (LOCC) \cite{BoseQGExp,BoseLocal} or, more generally, as a local classical-information  mediator \cite{VedralQGExp,PhysRevD.98.046001,marletto2019answers}, which can never create entanglement. However, it is in theory possible that CG could cause two spatially separated quantum matter systems  to  directly couple with one another,  invalidating the LOCC and classical-information mediator arguments and leading to entanglement generation in experiments  \cite{BoseQGExp,BoseLocal,VedralQGExp,PhysRevD.98.046001}. For instance, such direct CG interactions could be due to non-local effects associated with CG \cite{BoseQGExp,BoseLocal,VedralQGExp,PhysRevD.98.046001}, or even quasi-local CG effects, such as tunnelling between  two   quantum matter systems, which we consider in Section \ref{sec:DirectInts}.

\begin{figure}
	\begin{center}
		\hspace{0.35cm}
		\subfigure{%
			\put(-13,100){(a)}
			\label{fig:EGSphere}
			\includegraphics[width=0.175\textwidth]{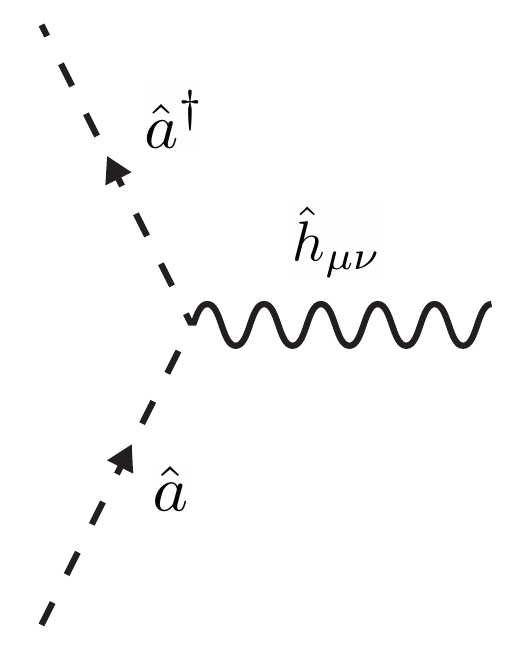}            
		}%
		\hspace{0.75cm}
		\subfigure{%
			\put(-13,100){(b)}
			\label{fig:EGRateOfChangeSphere}
			\includegraphics[width=0.175\textwidth]{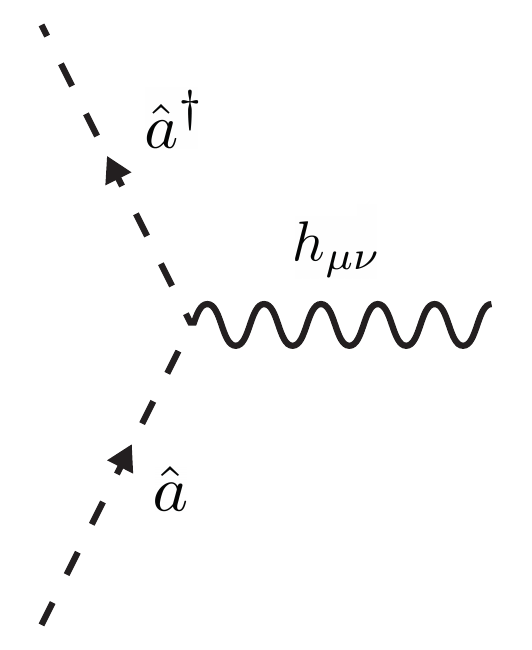}
		}
	\end{center}
	\vspace{-0.5cm}
	\caption{(a) Basic Feynman diagram for matter interacting with QG where matter emits a graviton, which is associated with $\hat{h}_{\mu \nu}$. For simplicity, we  represent matter by a real scalar field such that $\hat{a}\da$ and $\hat{a}$ are the creation and annihilation operators of matter. The interaction is then associated with three quantum operators and, therefore, can induce non-Gaussianity. (b) We can illustrate the analogous interaction between matter and  classical gravity with a similar diagram except that now the gravitational leg represents a classical gravitational wave $h_{\mu \nu}$ rather than a graviton. Since this CG interaction is  associated with just two quantum operators $\hat{a}\da$ and $\hat{a}$, it cannot, in contrast to the QG interaction, induce non-Gaussianity. Note that, although these diagrams represent weak-field, perturbative gravitational interactions, the fact that CG cannot create non-Gaussianity also applies to the strong-field, non-perturbative regime of gravity, as shown in Section \ref{sec:TheTheory}.}
	\label{fig:Feynm} 
\end{figure}

Here, we take a radically different approach to testing QG with QIS. Rather than concentrating on how QG can act as a quantum-information mediator in comparison to  a classical communication channel \cite{BoseQGExp,BoseLocal} or  classical-information mediator \cite{VedralQGExp,PhysRevD.98.046001}, we consider how just the simple process of adding a hat to classical gravitational degrees of freedom, i.e.\ turning them into quantum operators, results in a theory that, in contrast to its classical counterpart, can create non-Gaussianity in the quantum field of matter. Non-Gaussianity is a key resource in  continuous-variable  QIS (CVQIS), where quantum information is encoded in  degrees of freedom  with a continuous spectrum. For example, it is necessary in order to perform  universal quantum computation \cite{PhysRevLett.82.1784,PhysRevA.68.042319}. In fact, the reason that, in the exclusion of all other quantum interactions, only a quantum rather than classical theory of gravity can create non-Gaussianity in the quantum field state of matter is for the same reason that non-Gaussianity is  required for  universal quantum computation:  non-Gaussianity is created by processes that are non-quadratic in  quantum operators \cite{Sorkin_2014,QFsQIs,Berges2018,InfModes}, and   only QG, compared to CG, can contain such processes. Although  our argument applies to both the weak- and strong-field regimes of gravity, it is perhaps most intuitively understood from a perturbative weak-field perspective. In this case, the  simplest interaction  between matter and gravity in QG is  where matter creates a graviton. As illustrated in Figure \ref{fig:Feynm}, the corresponding Feynman diagram contains three quantum operators and, therefore, induces non-Gaussianity. On the other hand, in CG, we would remove the hat of the gravitational degrees of freedom, leaving a quadratic Hamiltonian, which preserves Gaussianity \cite{GQI}.



In comparison to entanglement, since our non-Gaussianity indicator is not based on  LOCC or classical-information mediator arguments, this indicator of QG is not reliant on the non-existence of direct CG interactions,  which we illustrate in Section \ref{sec:DirectInts}.  Therefore, as long as we are  working in an experimental situation where non-gravitational quantum interactions can be neglected (just as in tests based on entanglement \cite{BoseQGExp,BoseLocal,VedralQGExp,PhysRevD.98.046001}),  non-Gaussianity can be used as a signature of QG in experimental tests without the need for the additional assumption of there being no direct CG interactions  \cite{BoseQGExp,BoseLocal,VedralQGExp,PhysRevD.98.046001}.  A further advantage of a non-Gaussianity  signature is that a single system can be non-Gaussian, allowing for tests of QG that are based on just a single rather than multi-partite system. We illustrate this with a  table-top test of QG that uses a single Bose-Einstein condensate (BEC) in a single location. 

In addition to being just a single quantum system without any spatial superposition, our experimental proposal also  uses a type of quantum technology (BECs) for which certain massive quantum states have already been created \cite{kovachy2015quantum,Lange416,Kunkel413,Fadel409,Linnemann_2017}. This is in  contrast to  proposals based on opto-mechanical setups where massive non-classical states have yet to be generated. BECs also offer a contrasting method to distinguish the QG signal from electromagnetic noise. As with previous table-top proposals, it is vital that we can attribute the searched for QG effect from the analogous effect that is generated through electromagnetic interactions. For the BMV this effect is entanglement, which electromagnetic  as well as gravitational interactions will naturally generate, whereas, in the test proposed here this is non-Gaussianity, which electromagnetic interactions would also naturally generate since they are fundamentally quantum interactions.
	
To isolate the gravitational non-Gaussian signal from an electromagnetic one, we use the fact that  only the  electromagnetic interaction can be screened and, in particular, that BECs generically have Feshbach resonances. Electromagnetism can be screened since it has both positive and negative charges, whereas gravity is universal, coupling to all forms of energy in the same way. In BECs, this is immediately apparent since the atoms have zero overall electromagnetic charge, resulting in them naturally only interacting through  van der Waals and, in most cases, magnetic dipole-dipole  interactions (MDDIs) at very low temperatures. This allows for the use of an extraordinary  property of BEC and cold atom experiments to distinguish the electromagnetic and  gravitational effects. This property is the presence of optical and magnetic Feshbach resonances that are used in BEC experiments to control the strength of the   electromagnetic interactions between the atoms by the application of an external magnetic or optical field. This has become a vital tool of BEC experiments that has facilitated numerous explorations of fundamental physics \cite{RevModPhys.82.1225,PitaevskiiBook}.
	
By applying a magnetic or optical field to the BEC, we can  in principle set the overall strength of the relevant electromagnetic interactions to zero without affecting the strength of the gravitational interactions \cite{RevModPhys.82.1225,PitaevskiiBook}. Then, a non-Gaussian signal can be attributed to only  gravitational interactions. This method contrasts with that used in opto-mechanical proposals where the distance between micro-objects is increased to a level where gravitational interactions are greater than the electromagnetic van der Waals interactions. In that case, both the   electromagnetic and gravitational interactions are suppressed by increasing the distance, whereas applying an external magnetic or optical field to a BEC only affects the former.

\section{Non-Gaussianity as a signature of quantum gravity} \label{sec:TheTheory}

Consider a free, real scalar  quantum field $\hat{\phi}$. The Hamiltonian of this system can  be written as a collection of quantum simple harmonic oscillators: $\hat{H} = \sum_k \hbar \omega_k [ \hat{a}\da_k \hat{a}_k + 1/2]$, where $\hat{a}\da_k$ and $\hat{a}_k$ are creation and annihilation operators of mode $k$; $\omega_k$ is the angular  frequency; and we have assumed  a discrete mode spectrum  for simplicity \cite{peskin1995introduction}. For each oscillator we can associate position and momentum-like operators, $\hat{x}_k := \hat{a}_k + \hat{a}_k\da$ and $\hat{p}_k := i(\hat{a}\da_k - \hat{a})$, known as quadrature operators,  which are observables with a continuous eigenspectra: $\hat{x}_k |x \rangle_k = x_k | x\rangle_k$ and $\hat{p}_k |p \rangle_k = p_k | p\rangle_k$. The quadrature eigenvalues, $x_k$ and $p_k$, can be used as continuous variables to describe the entire quantum field system, and we can view this as a continuous  phase space on which we encode our quantum information \cite{GQI}. This approach to encoding quantum information  can also be straightforwardly extended to general bosonic and fermionic quantum fields \cite{Corney_2005,FermionGauss,NonEqQFT,greplova2013quantum,hackl2018aspects,dangniam2018quantum}.  

Rather than describing this scalar field system using a density operator $\hat{\rho}$, an equivalent representation is provided by the Wigner function \cite{PhysRev.40.749}. This  is a quasi-probability distribution defined over  phase space, analogous to probability distributions  used in classical statistical mechanics. For example, for a single mode of a scalar field, the Wigner function can be obtained through \cite{hillerym106connel}:
\begin{align}
W_{\hat{\rho}}(x, p)=\frac{1}{2\pi } \int dy\, e^{-i y p }\langle x+y|\hat{\rho}| x-y\rangle.
\end{align}
$W_{\hat{\rho}}(x, p)$ is a quasi-probability distribution since, although it takes on real values and is normalized to unity, it can also take on negative values. The states for which the above Wigner function takes on negative values, therefore, have no classical counterpart, and are considered to be highly non-classical states \cite{Kenfack_2004}. 

For the scalar field, the only states that have negative Wigner functions are non-Gaussian states, such as  Fock states or Schr\"{o}dinger cat states\footnotetext[3]{For pure bosonic states, the divide between states with positive and negative Wigner functions exactly coincides with the divide between states with   Gaussian and non-Gaussian Wigner functions. In contrast, for general mixed states, although only non-Gaussian states can have negative Wigner functions, it is also possible for a non-Gaussian state to have a positive Wigner function.} \cite{Note3}. Gaussian states,  on the other hand, such as coherent states, squeezed states and thermal states, have only positive Wigner functions \cite{HUDSON1974249,GQI}.  Here, we define a Gaussian state as a state that is fully characterized by the first and second moments of the quadrature operators, or, equivalently, by the one and two-point correlation functions of the quantum field \cite{GQI,Sorkin_2014,Berges2018}. We use this definition for all types of fields, bosonic and fermionic \cite{GQI,Corney_2005,FermionGauss,NonEqQFT,greplova2013quantum,hackl2018aspects,dangniam2018quantum}. For our scalar field $\hat{\phi}$, the Wigner function of such a state  is also a Gaussian distribution \footnotetext[44]{For  fermionic particles and fields, this may not necessarily be the case \cite{PhysRevA.18.1250,HAKIM1982230,ELZE1986402,VASAK1987462,PhysRevD.44.1825,doi:10.1142/7881,refId0,hackl2018aspects}. It has been also found that Hudson's theorem stating that the only wave functions that have negative Wigner functions are non-Gaussian does not apply to Dirac fermions \cite{PhysRevA.90.034102}. Alternative quasi-probability distributions for Dirac fields that obey this theorem are under investigation \cite{dangniam2018quantum}.}\cite{GQI,Note44}.  

The classification of Gaussian and non-Gaussian states is very important in CVQIS. For example, universal quantum computation with pure states is only possible with non-Gaussian  states or operations \cite{PhysRevLett.82.1784,PhysRevA.68.042319}, while  Gaussian states and operations can be efficiently simulated on a classical computer \cite{PhysRevLett.88.097904,PhysRevLett.89.137903,PhysRevLett.89.137904,PhysRevA.66.032316}. Furthermore, non-Gaussian states or operations are required for   violation of  Bell inequalities \cite{PhysRevA.58.4345,PhysRevLett.82.2009,PhysRevA.66.044309,PhysRevLett.88.040406,PhysRevLett.93.020401,PhysRevA.72.042105,PhysRevA.71.022105,Ferraro_2005}.   These, and additional examples, such as implementing entanglement distillation \cite{PhysRevA.66.032316},  have led to non-Gaussianity being classified as a CVQIS resource for which measures and witnesses have been derived \cite{PhysRevA.76.042327,PhysRevA.97.052317,PhysRevA.78.060303,PhysRevA.98.052350,PhysRevA.82.052341,PhysRevA.88.012322,EffQuantNonGauss,PhysRevA.87.033839,PhysRevA.90.013810,PhysRevA.87.062104}, just as for entanglement. 

Given the significance of Gaussian and non-Gaussian states in CVQIS, it is important to distinguish the type of Hamiltonians that can create such states: a Hamiltonian that is at most quadratic in quadratures, or equivalently in annihilation and creation operators, can only ever map a Gaussian state to another Gaussian state \cite{Schumaker1986317,Cheng_1988,LITTLEJOHN1986193,PhysRev.162.1256,NonEqQFT,LieAlgGauss}, which holds for both bosonic and fermionic \cite{Corney_2005,greplova2013quantum,hackl2018aspects,dangniam2018quantum} fields. That is, the Hamiltonian must be of the form:

\begin{align} \label{eq:Hquad}
\hat{H} =  \sum_k \bs{\lambda}_k(t) \hat{\bs{x}}_k  + \sum_{k,l} \hat{\bs{x}}^T_k \bs{\mu}_{kl} (t) \hat{\bs{x}}_l,  
\end{align}
where $\hat{\bs{x}}^T_k := (\hat{x}_k, \hat{p}_k)$, and $\bs{\lambda}_k(t)$ and $\bs{\mu}_{kl} (t)$ are $2 \times 1$ and $2 \times 2$ real-valued matrices of arbitrary functions of time. Although we have  assumed a discrete, finite mode spectrum here for simplicity,  the extension to infinite and continuous modes is straightforward \cite{Sorkin_2014,QFsQIs,Berges2018,InfModes}. 

The  Hamiltonian \eqref{eq:Hquad}  preserves Gaussianity since it is associated with a general Bogoliubov transformation, which is a linear transformation of the  quadratures (and, therefore,  phase space) that  preserves their   commutation relations, or anti-commutation relations for fermionic fields. Any other Hamiltonian, i.e.\ one that is  not linear or quadratic in quantum operators, will in general create non-Gaussianity  \cite{GQI,PhysRevLett.82.1784,PhysRevA.68.042319}.

Note that a free quantum field has a Hamiltonian that is of the form \eqref{eq:Hquad} since it only contains the kinetic and mass terms, and so is necessarily quadratic in the field. For example, the free Hamiltonian for a real scalar quantum field $\hat{\phi}$ is \cite{peskin1995introduction}:
\begin{align} \label{eq:Hphi}
\hat{H} &= \frac{1}{2} \int d^3 \bs{r} \left[(\partial_t \hat{\phi})^2 + (\nabla \hat{\phi})^2 + m^2 \hat{\phi}^2 \right]
\end{align}
where $m$ is the  mass of the field. Expanding the field in annihilation and creation operators $\hat{\phi} = \sum_k [u_k (t) \hat{a}_k + v(t) \hat{a}_k\da]$, results in a Hamiltonian of the form \eqref{eq:Hquad} \cite{peskin1995introduction}.

Now consider interacting this  quantum field with a classical entity $\mathcal{G}$, which could depend on space and time. Taking the classical interaction to not induce quantum self-interactions of $\hat{\phi}$, then $\mathcal{G}$ and $\hat{\phi}$ can only interact through  Hamiltonian terms that are linear or quadratic in $\hat{\phi}$\footnotetext[25]{We do not include terms that are neither linear nor quadratic in $\hat{\phi}$, such as  $\hat{\phi}^3 \mathcal{G}$, since we consider these as inducing quantum self-interactions of $\hat{\phi}$.  Later, when we consider $\mathcal{G}$ to be associated with gravitational interactions, such terms will turn out to be excluded since we will be working in a situation where all non-gravitational interactions can be ignored. Similarly, if we single out any of the Standard Model forces and take them to be classical, then these terms are  not  present.} \cite{Note25}. For example, the classical interaction could occur through a Hamiltonian term such as $(\nabla \hat{\phi})^2 f[\mathcal{G}]$, where $f$ is a real functional of $\mathcal{G}$. Then, expanding $\hat{\phi}$ in annihilation and creation operators,  we would still find a Hamiltonian that is of the form \eqref{eq:Hquad}, with $\mathcal{G}$ just absorbed into the time-dependent coupling constants. That is, the Hamiltonian of the classical interaction preserves Gaussianity, and this would  apply to a classical interaction with any type of quantum field, not just a real scalar field $\hat{\phi}$.

In contrast, if we quantize $\mathcal{G}$, such that we interact $\hat{\phi}$ (or any other type of quantum field) with a quantum entity, then it is possible for the resulting Hamiltonian to be higher order than  quadratic in quantum operators, and thus induce non-Gaussianity. Therefore,  any sign of the creation of non-Gaussianity in the state of a  quantum field would be evidence of a  quantum  interaction.

Due to the universal coupling of gravity,  we can apply this argument to determine whether gravity obeys a  quantum or classical theory. In this case, if we are working in a situation where all other  quantum interactions can be neglected,  the matter Hamiltonian  contains only the kinetic and mass terms of the matter quantum field, to which gravity couples. If there were terms that were neither linear nor quadratic in the quantum matter field, and which thus induce quantum self-interactions of matter, then these would have to be associated with a non-gravitational force since these terms must also be present in flat space. Therefore, as we are assuming a situation where all interactions other than gravity can be ignored, these terms are not present.

 For example, if, for simplicity, we used a real scalar field $\hat{\phi}$ to describe matter and ignored a possible quadratic Ricci scalar coupling term\footnotetext[38]{By `matter', we mean the leptons and quarks of the Standard Model, although we could also include other quantum entities, such as the electromagnetic field. The leptons and quarks are  described by spin-1/2 quantum fields, and we have  used a simple real scalar field $\hat{\phi}$ (for which particles are their own anti-particles) to describe matter only to illustrate our argument. The argument also applies to spin-1/2 fields.} \cite{Note38}, then the Hamiltonian of CG would be \eqref{eq:Hphi} but with $\sqrt{\mathfrak{g}}$ multiplying each term, where $\mathfrak{g}$ is the determinant of the spatial metric \cite{Thiemann_1998,rovelli_2004,thiemann2008modern}\cite{Note43}.\footnotetext[43]{Also see Appendix \ref{app:GRAction} for a brief summary.}  This Hamiltonian would preserve Gaussianity. In contrast, in QG there must be an operator associated with the gravitational field, which would result in Gaussianity no longer being preserved. For instance, in loop QG, $\mathfrak{g}$ would be quantized in this example \cite{Thiemann_1998,rovelli_2004,thiemann2008modern}\cite{Note43}, and in linearized QG, we would perturb the gravitational metric around a classical spacetime background metric and quantize only the perturbation \cite{Feynman:1963ax,PhysRev.162.1195,Gupta:1952zz}\footnotetext[39]{See Appendix \ref{app:NewtLimitQG} for the Hamiltonians of weak-field   gravity.}\cite{Note39}. Similarly, in the non-relativistic Newtonian limit, only the temporal component of the perturbed metric would be used, which is quantized and associated with the Newtonian gravitational potential $\Phi$ \cite{Note45}.

 In summary, creation of non-Gaussianity would provide evidence for a quantum theory of gravity. In fact, since all known  fundamental interactions with matter, such as electromagnetism, have interaction Hamiltonians with  terms that are quadratic in matter fields\footnotetext[22]{{By matter interactions, we mean all interactions of the quarks and leptons, i.e.\ the electroweak force, the strong force and the gravitational force.  We do not take, for example, the Higgs to be matter.}} \cite{peskin1995introduction,rovelli_2004}, non-Gaussianity  could also be used to evidence that these  are indeed quantum interactions \cite{Note22}.

\section{Testing quantum gravity with a single quantum system} \label{sec:Test}

\begin{figure}
 	\hspace{-2.0cm}
 	\subfigure{%
 		\put(-13,90){(a)}
 		\includegraphics[width=0.2\textwidth]{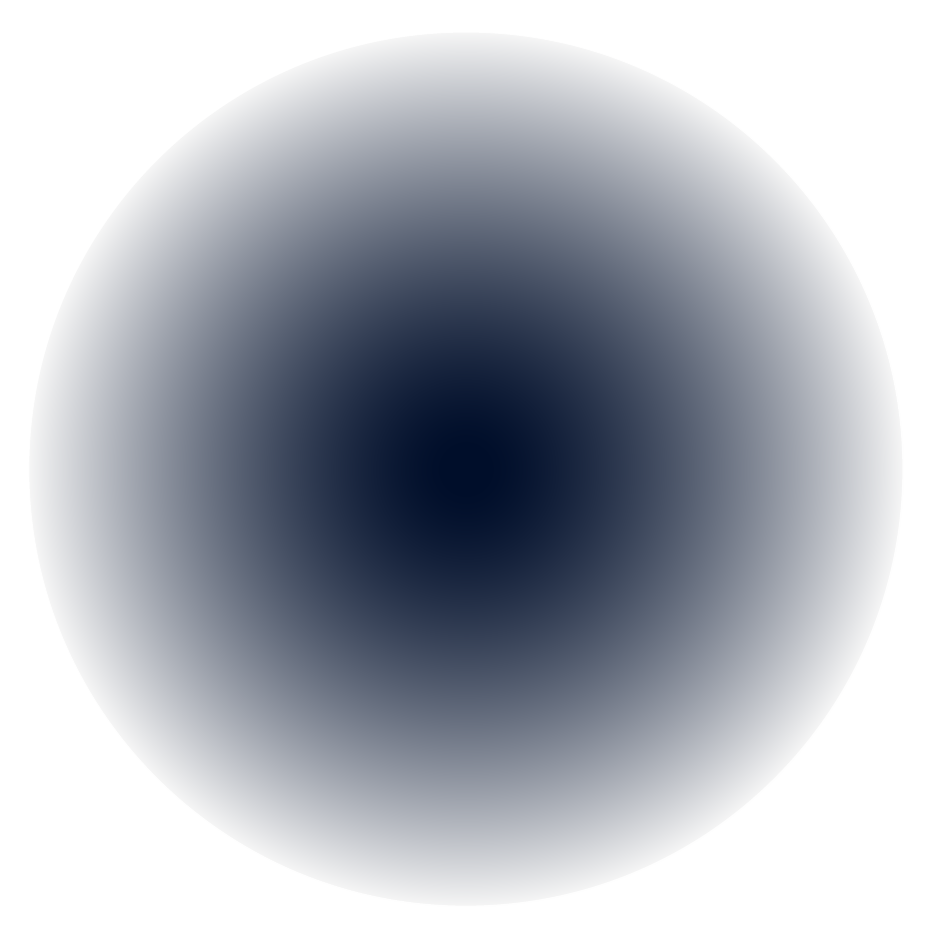}            
 	}%
 	\\
 	\vspace{-0.1cm}
 	\hspace{-0.3cm}
 	\subfigure{%
 		\put(-12.5,60){(b)}
 		\includegraphics[width=0.3\textwidth]{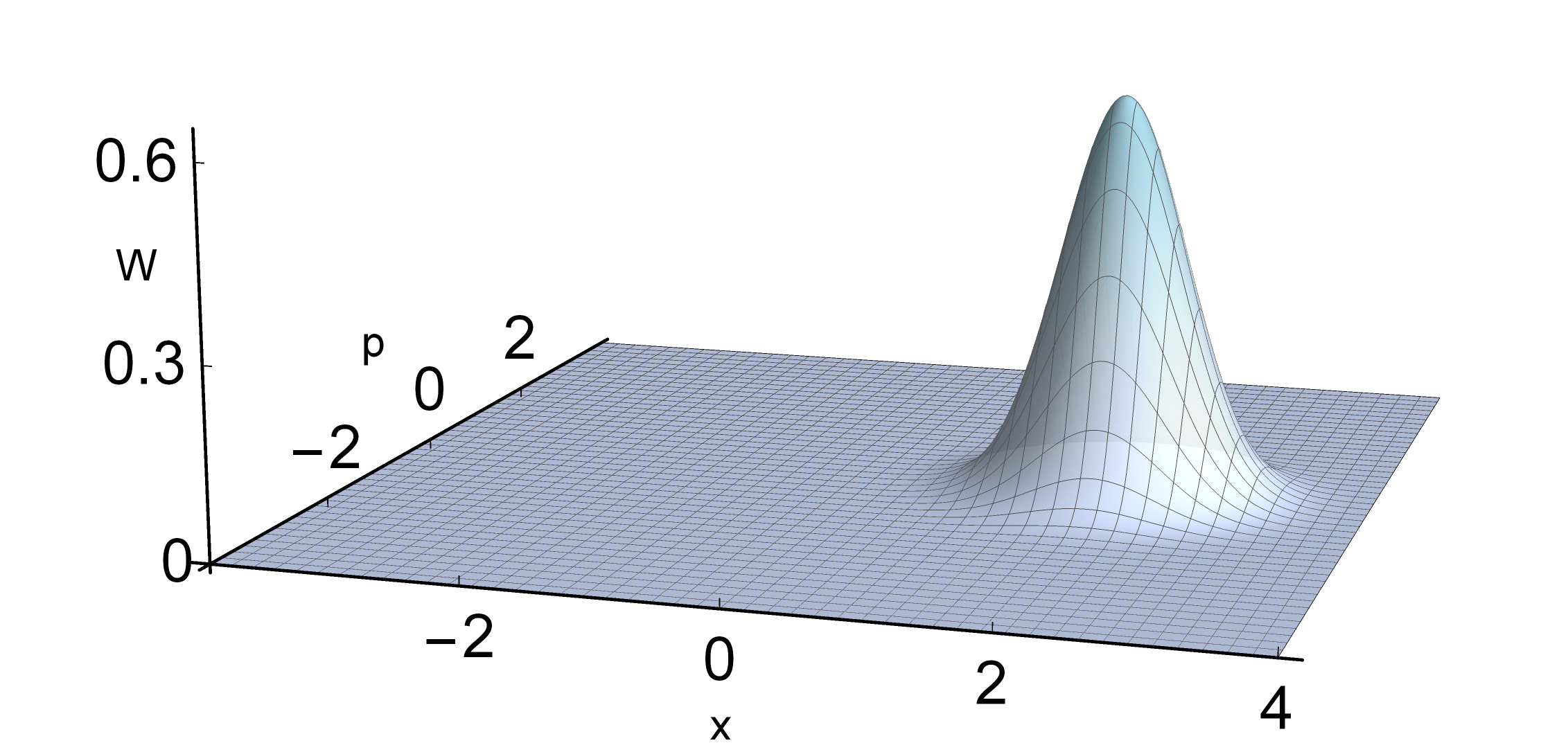}
 	}
 	\\
 	\vspace{0.1cm}
 	\hspace{0.1cm}
 	\subfigure{%
 		\put(-16.5,60){(c)}
 		\includegraphics[width=0.3\textwidth]{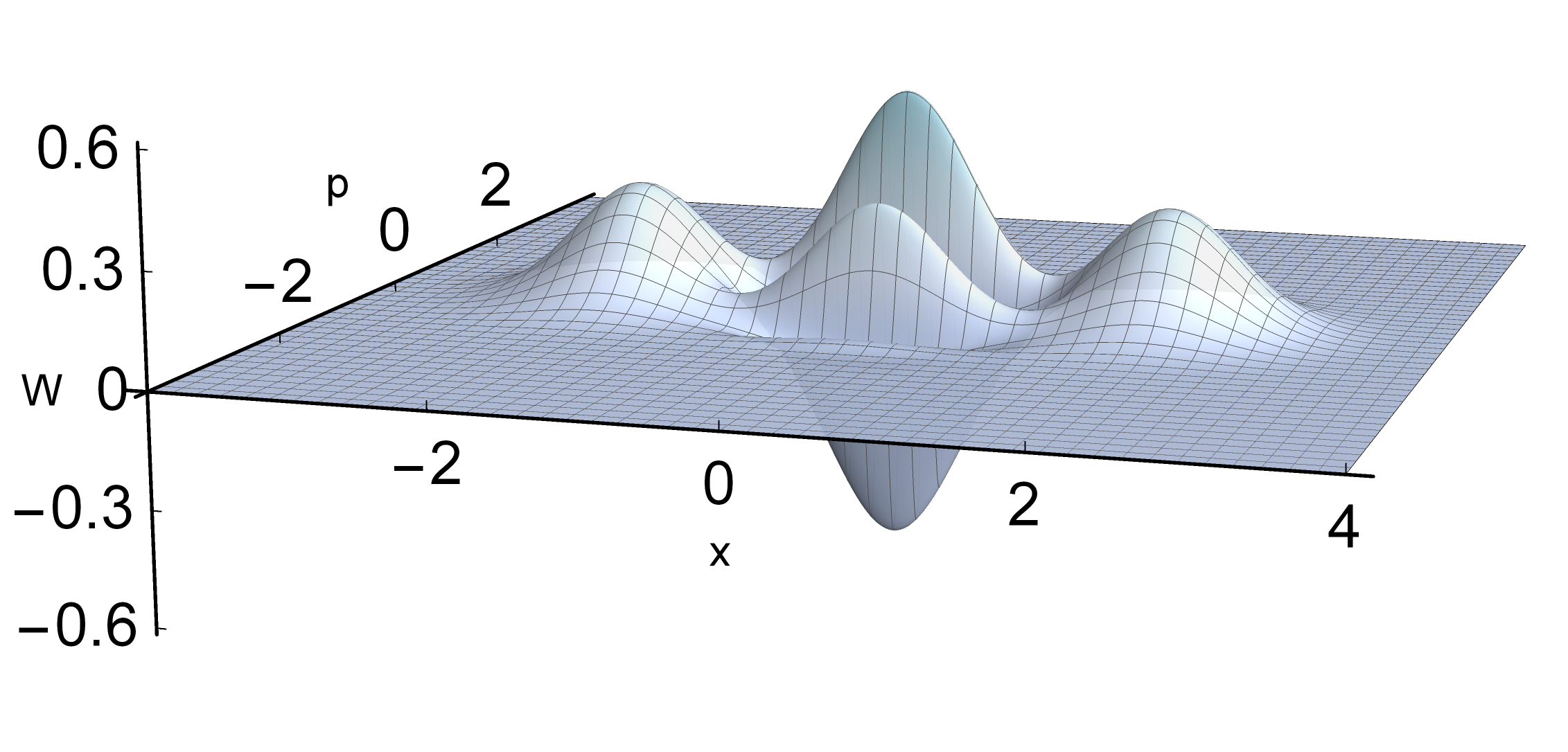}
 	}
 	\caption{Illustration of the proposal outlined in Section \ref{sec:Test} for a table-top test of QG. (a) A single spherical BEC of $10^9$ atoms and radius $R=200\unit{\mu m}$ is left to self-interact gravitationally for around $t = 2\unit{s}$. If QG acts, non-Gaussianity is induced in its quantum state, whereas, if CG acts, Gaussianity is preserved. Each atom is equally delocalized across the extent of the BEC but, since the BEC is in a spherical harmonic trap, the density of the BEC is greatest at its centre and drops off to zero asymptomatically, as illustrated. The BEC is initially in a Gaussian state, and (b) displays  its Wigner function $W$. Here, for simplicity, a coherent state $|\alpha \rangle$ is assumed, but this can also be squeezed to improve the signal-to-noise ratio, as discussed in Section \ref{sec:Test}. If QG acts, and the interaction time were long enough, then gravity could even force the coherent state to a Yurke-Stoler cat state $(|\alpha \rangle + i |-\alpha \rangle)/\sqrt{2}$. The Wigner function for such a non-Gaussian state is illustrated in (c). If, however, CG acts, then the state will remain Gaussian. In the non-relativistic CG limit, the state will in fact remain a coherent state, whereas, relativistic CG effects would, in principle, squeeze the state but keep it Gaussian. In practice, the interaction time would not be long enough for such a dramatic effect as a coherent state changing to a Yurke-Stoler state, and instead smaller deviations from a Gaussian distribution are looked for (see Section \ref{sec:Test}). Note that $\alpha = 2$ is used in the plots, whereas, in practice, the BEC will have an amplitude of around $|\alpha| = 10^{4.5}$, the square-root of the number of atoms.}     
 	\label{fig:BEC} 
 \end{figure}

We now consider a  table-top test of QG that uses our non-Gaussianity witness (Figure \ref{fig:BEC}). This experiment is based on a single BEC that is in a single location, and is an experiment to which an entanglement witness of QG could not be applied.

\subsection{Non-Gaussianity as a signature of quantum gravity in a Bose-Einstein condensate}

A Bose gas can be described by a non-relativistic scalar quantum field $\hat{\Psi}(\bs{r})$, which creates an atom at position $\bs{r}$ \cite{PitaevskiiBook}. Assuming that we are working at low enough temperatures such that the ground-state is macroscopically occupied, we neglect the thermal component of the gas and take $\hat{\Psi} (\bs{r})  \approx \psi (\bs{r}) \hat{a}$, where $\psi(\bs{r})$ is the wave-function of a condensed atom, and $\hat{a}$ is the annihilation operator for the condensate \cite{PitaevskiiBook}. The identical atoms are then all in the same state, have the same wavefunction, and are equally delocalized across the BEC.

These atoms will  interact gravitationally with each other, and since this is a non-relativistic system, it is appropriate to take the non-relativistic (Newtonian) approximation of gravity, which all gravitational theories must limit to. The  fully classical interaction Hamiltonian  for  Newtonian gravity is:
\begin{align} \label{eq:HintNG}
H_{int} = \frac{1}{2} \int d^3 \bs{r} \rho (\bs{r}) \Phi (\bs{r}),
\end{align}
where $\Phi(\bs{r})$ is the classical Newtonian potential and $\rho(\bs{r})$ is the classical mass density. If gravity obeys quantum theory then we must quantize both $\rho (\bs{r}) $ and $\Phi(\bs{r})$, whereas, if we have CG then we only quantize the former. Since $\hat{\rho} (\bs{r}) = m \hat{\Psi}\da (\bs{r}) \hat{\Psi}\da (\bs{r}) $ for a BEC, this results in the respective QG and CG interaction Hamiltonians:
\begin{align} \label{eq:HintQGNewt}
\hat{H}_{QG} &= \frac{1}{2} m \int d^3 \bs{r} :\hat{\Psi}\da (\bs{r}) \hat{\Psi} (\bs{r}) \hat{\Phi} (\bs{r}):\\ \label{eq:HintCGNewt}
\hat{H}_{CG} &= m \int d^3 \bs{r} \hat{\Psi}\da (\bs{r})  \hat{\Psi}(\bs{r}) \Phi[\varPsi] (t,\bs{r}),
\end{align} 
where $::$ refers to normal ordering, $m$ is the mass of the atoms, and we have made explicit that the classical potential $ \Phi$ can be a functional of the quantum state $\varPsi$ of the BEC, for which we have dropped a factor of $1/2$ \cite{Mielnik1974}. Solving the quantized version of Poisson's equation, we have:
 \begin{align}
 \hat{\Phi}(\bs{r}) = - G m \int d^3 \bs{r'}\, \frac{\hat{\Psi}\da(\bs{r'}) \hat{\Psi}(\bs{r'})}{  |\bs{r} - \bs{r'}|},
 \end{align} 
where $G$  is the gravitational constant. In contrast, depending on the chosen CG theory, $\Phi$ is  a certain quantum average of this expression (for example, in the CG Schr\"{o}dinger-Newton theory, which is the Newtonian limit of the semi-classical CG theory, $\Phi = \langle \hat{\Phi} \rangle$ \cite{DIOSI1984199,PenroseSNEquationsAndComputation,RoadToReality} \footnotetext[42]{{Note that  Schr\"{o}dinger-Newton  can refer to the  CG theory where gravity is \emph{fundamentally} sourced by the expectation value of quantum matter density \cite{PenroseSNEquationsAndComputation,Bahrami_2014}, as well as to a limit of QG where a Hartree approximation for the mutual quantum gravitational interaction in a system of many particles is used. These two descriptions are fundamentally different, this being most clear in the single particle limit where the latter description cannot be used but where the CG theory still applies and describes a single particle wavefunction self-interacting gravitationally (which cannot happen in QG) \cite{Bahrami_2014}.  	Here, by the Schr\"{o}dinger-Newton equations, we refer to the fundamental CG theory rather than the QG approximation.}}\cite{Note42}).

 Using $\hat{\Psi}(\bs{r}) = \psi(\bs{r}) \hat{a}$, the above interaction Hamiltonians for the BEC reduce to: 
\begin{align}\label{eq:QGBEC}
\hat{H}_{QG} &=   \frac{1}{2} \lambda_{QG} \hat{a}^{\dagger} \hat{a}^{\dagger} \hat{a} \hat{a},\\\label{eq:CGBEC}
\hat{H}_{CG} &= \lambda_{CG} [\varPsi] \hat{a}^{\dagger} \hat{a},
\end{align}
where \cite{Note7,Note23}:
\begin{align} \label{eq:lambdaQG}
\lambda_{QG} &:=-  G m^2 \int d^{3} \bs{r}\, d^{3} \bs{r'} \frac{|\psi(\bs{r'})|^2 |\psi(\bs{r})|^2 }{\left|\bs{r}-\bs{r'}\right|},\\ \label{eq:lambdaCG}
\lambda_{CG}[\varPsi] (t) &:=G m \int d^{3} \bs{r}  |\psi(\bs{r})|^2 \Phi[\varPsi] (t,\bs{r}).
\end{align}
\footnotetext[7]{{Note that, although $\lambda_{QG}$ and $\lambda_{CG}$ may look very similar, they can take on very different values. For example, in the  Schr\"{o}dinger-Newton equations, $\lambda_{CG} = N \lambda_{QG}$, where $N$ is the number of atoms of the BEC (see Appendix \ref{app:CG}).}}\footnotetext[23]{Note that, in the Newtonian limit, the QG Hamiltonian \eqref{eq:QGBEC} appears as an effective quantum self-interaction of matter. In contrast, the CG Hamiltonian cannot induce quantum self-interactions of matter \eqref{eq:CGBEC}. This is because, in the absence of all other interactions, the CG Hamiltonian  only contains the gravitational field coupled to the kinetic and mass terms of matter, which are quadratic in the matter field.}  The QG interaction Hamiltonian \eqref{eq:QGBEC} can also be derived as the non-relativistic limit of linearized QG where we  consider the four-point Feynman diagram with a single virtual graviton propagator, and then effectively integrate out  gravitational degrees of freedom \cite{PhysRevLett.72.2996,PhysRevD.50.3874,AKHUNDOV199716,PhysRevD.67.084033,burgess2004quantum,maggiore2008gravitational}\footnotetext[45]{Also, see Appendix \ref{app:NewtLimitQG} for a derivation of Newtonian gravity from GR.}\cite{Note45}. All QG theories must limit to the above quantum version of Newtonian gravity and so \eqref{eq:QGBEC} is the Hamiltonian for general QG self-interactions of a  BEC. Likewise, all CG theories must limit to \eqref{eq:CGBEC} for CG self-interactions of a BEC in the Newtonian limit, such that \eqref{eq:QGBEC}-\eqref{eq:CGBEC} are not dependent on a specific model of CG or QG.  Similar Hamiltonians have been derived  using cold atoms in a double-well potential \cite{2018arXiv181010202H}.

From the Hamiltonians \eqref{eq:QGBEC}-\eqref{eq:CGBEC}, we can see that, as long as all other quantum interactions can be neglected (see Section \ref{sec:DistOtherQMInts}), only the  QG Hamiltonian \eqref{eq:QGBEC}  can induce non-Gaussianity in the quantum state of the BEC field, with the CG Hamiltonian \eqref{eq:CGBEC} preserving Gaussianity since it is quadratic in quantum operators. Therefore, any sign of non-Gaussianity being created in the BEC would be evidence of QG. Note that entanglement cannot be used as a witness here since this is just a single-mode system\footnotetext[10]{{Entanglement has been observed in split BECs \cite{Lange416,Kunkel413,Fadel409}. In these experiments the initial state was of two BECs in two hyperfine levels. Although the initial BECs do not appear to be entangled in the second quantization picture, they do look entangled in the first quantization picture (this is often referred to as ``particle'' entanglement). In contrast, since we only have a single BEC here in a single location, there is no entanglement in either quantization picture. In the second quantization picture, a single ket is used to describe the system, such as  $|\alpha\rangle$ for a coherent state. In the first quantization picture, in the limit of absolute zero, and fixing the particle number as $N$, the state of the system is described by the many-body wavefunction  $\varPhi (\mathbf{r}_{1}, \mathbf{r}_{2}, \ldots \mathbf{r}_{N})=\psi_0(\mathbf{r}_{1}) \psi_0(\mathbf{r}_{2}) \ldots \psi_0(\mathbf{r}_{N})$ where $\psi_0$ are the identical wavefunctions for each atom.}}\footnotetext[8]{As indicated in \eqref{eq:lambdaCG}, CG theories can be non-linear in the evolution of the state vector \cite{Mielnik1974}. However, even if it is non-linear,  it is still  a Gaussian process since a CG theory must be quadratic in matter fields: the non-linearity  means that, although we know a Gaussian state will remain a Gaussian state, we may not be able to analytically determine the specific evolution (see Appendix \ref{app:CG} for more detail).} \cite{Note10,Note8}. In fact, the QG Hamiltonian is analogous to the Kerr interaction, which induces non-Gaussianity in quantum optics \cite{KnightBook}.

\subsection{Measurement scheme}

As shown above, measuring creation of non-Gaussianity in the BEC would provide evidence of QG. In order to detect  non-Gaussianity, we consider measurements of high-order cumulants \cite{EffQuantNonGauss}. For a Gaussian distribution, all cumulants higher than second order  vanish and, therefore,  a non-zero value of such cumulants is a signature of non-Gaussianity. Here we concentrate on the  fourth-order cumulant $\kappa_4$, since $\kappa_3$ is also zero for a symmetric non-Gaussian distribution. Defining a generalized quadrature as $\hat{q} (\varphi) = \hat{a} e^{-i \varphi} + \hat{a}\da e^{i \varphi}$, we have:
\begin{align}\label{eq:k4}
\kappa_4 := \langle\hat{q}^4\rangle - 4 \langle\hat{q}\rangle \langle \hat{q}^3\rangle -3\langle\hat{q}^2\rangle^2  + 12 \langle\hat{q}^2\rangle \langle\hat{q}\rangle^2 - 6 \langle\hat{q}\rangle^4.
\end{align}
In an experiment, only a finite sample can be used to estimate  $\kappa_4$ and we desire unbiased estimators, which  are the $k$ statistics: $\braket{k_n} = \kappa_n$ \cite{kendall1948advanced}. The  noise in the estimation of $\kappa_4$ is then the standard deviation of $k_4$\footnotetext[41]{See Appendix \ref{app:SNR} for the standard deviation of $k_4$.} \cite{Note41}, such that the signal-to-noise ratio (SNR) for the measurement is:
\begin{align} \label{eq:SNR}
\text{SNR} =    |\kappa_4| /\sqrt{ \mathrm{Var} (k_4)},
\end{align}
where, for a large number of independent measurements $\mathcal{M}$,  $\mathrm{Var} (k_4) \propto 1/ \mathcal{M}$. 

In order to make the SNR as large as possible, we use quantum metrology, where highly quantum states can improve the estimation of parameters that are not associated with  observables \cite{giovannetti2011advances}. This is also effectively    used in the BMV proposal where the initial quantum states are N00N-like states \cite{PhysRevA.40.2417,PhysRevLett.85.2733,2018arXiv181010202H,2018arXiv181209776Q}. However, rather than using a N00N state, here we consider a squeezed state, which is a Gaussian state that often provides similar performance in quantum metrology to N00N states but which is usually far less demanding to create \cite{QuantumLimitsOptics}. Assuming that QG acts, i.e.\ that the gravitational interaction has the Hamiltonian   of QG \eqref{eq:QGBEC}, and taking the limit that $\chi := |\lambda_{QG}| / \hbar$ is small and that the number of atoms $N$ of the BEC   is large, the SNR can be  of  order $ \chi t N^2 \sqrt{\mathcal{M}}$, where $t$ is the interaction time\footnotetext[33]{See Appendix \ref{app:SNR}.} \cite{Note33}. Assuming a weakly interacting BEC of mass $M$ in a spherical harmonic trap with frequency $\omega_0$, the BEC wavefunction is \cite{PitaevskiiBook}:
\begin{align}\label{eq:psi}
	\psi(\bs{r}) = \frac{1}{\pi^{3/4} R^{3/2}} e^{-r^2 / (2 R^2)},
\end{align}
where $ R := \sqrt{\hbar / (m \omega_0)}$ is the effective radius of the spherical BEC and $r := |\bs{r}|$. Using \eqref{eq:lambdaQG}, this results in \cite{Howl_2019}:
\begin{align} \label{eq:size}
\chi t N^2 \equiv \sqrt{\frac{2}{\pi}} \frac{G M^2 t}{\hbar R},
\end{align}
which is $t / \hbar$ times the gravitational self-energy of the BEC. Note that, with the replacement of $R$ with $d$, and neglecting the numerical factor, this expression is  the same  as the relative phase generated in the BMV proposal between the two microspheres  that are separated by the smallest possible distance $d$ that, when ignoring all other distances, leads to  an  entangled state \cite{BoseQGExp,BoseLocal,VedralQGExp,PhysRevD.98.046001}. It is demonstrated in the BMV proposal that a value of order one for this phase is achieved   when  $d = 200\unit{\mu m}$,  $t\approx 2\unit{s}$  and  $M = 10^{-14}\unit{kg}$ \cite{BoseQGExp}.  However, since the SNR here scales with $\sqrt{\mathcal{M}}$, we can lower the total mass required by increasing the number of measurements. For example, to achieve an SNR of 5 for a $^{133}\mathrm{Cs}$ BEC, we could use  $R = 200\unit{\mu m}$,  $t= 2\unit{s}$ and  $M = 10^{-15}\unit{kg}$ with around 40,000 measurements. Such a mass corresponds to around $4 \times 10^9$ atoms, which is only a little larger than what has been achieved so far: in 1998 a  $^{1}\mathrm{H}$ BEC was created with over $10^9$ atoms \cite{HydrogenBEC}, and in 2006 a $^{23}\mathrm{Na}$ BEC had over $10^8$ atoms \cite{LargeNaBEC}. However, the number of atoms required can be reduced  by  further increasing $\mathcal{M}$.

An experimental implementation of this scheme would be to use a spin-1 BEC where the $m_F = \pm 1$ states are prepared in large coherent states and then a magnetic field is used to drive spin-mixing collisions to generate a quadrature squeezed state in the $m_F = 0$ condensate.  In a spin-1 BEC, the interaction Hamiltonian is \cite{PhysRevLett.81.5257,PhysRevA.87.063635,PumpedUpSU11}:
\begin{align} \label{eq:Hspin1} \nonumber
\hat{H} =\hbar \kappa&\left[\hat{a}_{0}^{2} \hat{a}_{+}^{\dagger} \hat{a}_{-}^{\dagger}+\left(\hat{a}_{0}^{\dagger}\right)^{2} \hat{a}_{+} \hat{a}_{-}\right] \\
&\nonumber+\hbar \kappa\left(\hat{a}\da_0 \hat{a}_0 -\frac{1}{2}\right)\left(\hat{a}\da_{+} \hat{a}_{+}+\hat{a}\da_{-} \hat{a}_{-}\right)\\&+\hbar q\left(\hat{a}\da_{+} \hat{a}_{+}+\hat{a}\da_{-} \hat{a}_{-}\right),
\end{align}
where $\hat{a}_0$ is the annihilation operator of the $m_F = 0$ mode and $\hat{a}_{\pm}$ are the annihilation operators of the $m_F = \pm 1$ modes. By dynamically tuning $q$ with a magnetic field, the quadratic Zeeman shift (third term)
cancels collisional shifts due to s-wave scattering of the three modes (second term) \cite{2014PhRvL.113j3004M,PumpedUpSU11}.  Taking  the $m_F = \pm 1$ modes to be in large coherent states ($N_{\pm} \gg 1$) so that $\hat{a}_{\pm} \approx \sqrt{N_{\pm}}$, \eqref{eq:Hspin1}  then acts as effectively:
\begin{align}
\hat{H} =\hbar N \kappa\left[\hat{a}_{0}^{2} + \left(\hat{a}_{0}^{\dagger}\right)^2\right] ,
\end{align}
where $N := \sqrt{N_+ N_-}$, which results in a single-mode quadrature squeezed state for the $m_F = 0$ mode \footnotetext[13]{{Alternatively,  rather than using a Gaussian squeezed state, another highly non-classical state could be used, such as a Schr\"{o}dinger cat state consisting of a superposition of coherent states with different phases.  For example,  a Yurke-Stoler state \cite{PhysRevLett.57.13} $ (\ket{\alpha} + i \ket{-\alpha}) /\sqrt{2}$ could, in principle, be created in a BEC using a magnetic field to ramp up the electromagnetic interactions between the atoms before subsequently turning them off \cite{PitaevskiiBook,KnightBook}. However, since the Yurke-Stoler state is non-Gaussian, we would have to either consider the change in the value of $\kappa_4$, or a protocol  where we evolve a single condensate in a coherent state $\ket{\alpha}$ to a Yurke-Stoler state, and then apply the reverse process such that the system returns to $\ket{\alpha}$ if the CG Hamiltonian \eqref{eq:CGBEC} acts rather than QG.}}\cite{Note13}. Spin-squeezing experiments have already been performed in cold atoms and BECs \cite{Linnemann_2017}, where normally it is the $m_F=0$ mode that is taken to be the large coherent mode and then a two-mode squeezed state is created for the $m_F =\pm 1$ modes. 

After the system has evolved for a time $t$, the non-Gaussianity of the BEC field would then be measured. To achieve this, a homodyne or heterodyne scheme could be used \cite{kurtsiefer1997measurement,gross2011atomic,PhysRevA.77.023619,PhysRevLett.99.010401}, where moments up to fourth order are looked for in the intensity difference, providing a direct map for obtaining $\kappa_4$. Observing a non-zero value for $\kappa_3$, which only requires the third-order moment in homodyne detection, would be sufficient for detecting non-Gaussianity, and  the third-order correlation function of atoms due to electromagnetic self-interactions has already been measured in experiments \cite{Hodgman1046}.	However, $\kappa_3$ is predicted to be zero if the initial state of the BEC is a squeezed vacuum state, in which case $\kappa_4$ needs to be analysed. For $\kappa_4$, the techniques used in  \cite{Hodgman1046} could be extended to measure the fourth-order correlations to obtain $\kappa_4$ through homodyne detection \cite{KnightBook}.  This would require single-atom counting in a quantum gas with high efficiency on small length scales, and recent advances have opened up very promising approaches to single-atom counting (see Appendix \ref{app:SAC} for more detail).  Rather than performing  a homodyne or heterodyne measurement, another option would be to  determine the Wigner function of the BEC, either using full state tomography with projective measurements (see e.g.\  \cite{kurtsiefer1997measurement,riedel2010atom,Schmied_2011,mcconnell2015entanglement,McConnell_2016} for such measurements on cold atoms and BECs), or through `direct' measurement with weak measurements of  the position quadrature and  projective measurements of the momentum quadrature \cite{lundeen2011direct,PhysRevLett.108.070402,Das_2017,Thekkadath_2018} (this  has so far been achieved with photons \cite{Hosten787,lundeen2011direct,malik2014direct,salvail2013full,PhysRevLett.112.070405}, but could be extended to atoms \cite{lundeen2011direct}). 

\subsection{Distinguishing quantum gravity from the electromagnetic interaction} \label{sec:DistOtherQMInts}

So far we have discussed how the desired input state can be generated and how non-Gaussianity could be measured. Additionally, it is imperative that we ensure that all noise can be distinguished from the signal. An advantage to considering a non-Gaussian signal is that we can immediately neglect all processes generating Gaussian noise  since these will not affect the non-Gaussian measurement.  The largest contributing non-Gaussian noise would be expected to come from the electromagnetic interactions between the atoms of the BEC. A BEC is very dilute and the atoms are neutral overall, but there are still, in general, weak electromagnetic interactions between the atoms due to van der Waals and MDDIs. At the low temperatures at which BECs operate, the Hamiltonian for a BEC with electromagnetic interactions is \cite{PhysRevLett.102.090402,PhysRevA.86.043625,PhysRevA.88.043609,tan2018effects,PhysRevLett.101.190405}:
\begin{align} \nonumber 
	\hat{H} &= \int d^3 \bs{r} \bigg[ - \frac{\hbar^2}{2m} \hat{\Psi}\da (\bs{r}) \nabla^2 \hat{\Psi}(\bs{r}) + V_T(\bs{r}) \hat{\Psi}\da (\bs{r}) \hat{\Psi} (\bs{r}) \\ \nonumber  &~~+  \frac{1}{2} \int d^3 \bs{r'} \Big[ \hat{\Psi}\da (\bs{r})  \hat{\Psi}\da(\bs{r'})  \hat{\Psi} (\bs{r})  \hat{\Psi}(\bs{r'}) \Big(g_s \delta^{(3)}(\bs{r}- \bs{r'}) \\ \label{eq:H} &~~+ g_d \frac{1 - 3 \cos^2 \vartheta}{| \bs{r}- \bs{r'}|^3} \Big) \Big]\bigg],
\end{align}
where the first term is the kinetic part, $V_T(\bs{r}) = m \omega_0^2 r^2/ 2$ is the spherical trapping potential, $g_s := 4 \pi \hbar^2 a_s / m$ is the s-wave scattering coupling constant,  $g_d := \mu_0 \mu^2 / (4 \pi)$ parametrizes the strength of the MDDIs, and $\vartheta$ is the polar angle of $\bs{r}- \bs{r'}$, with $a_s$ the s-wave scattering length, $\mu$ the magnetic moment of the atom and $\mu_0$ the permeability of free space. Using $\hat{\Psi}(\bs{r}) = \psi(\bs{r}) \hat{a}$ and \eqref{eq:psi},  the above Hamiltonian reduces to:
\begin{align} \label{eq:HBECFree}
	\hat{H} &= \hbar \omega \hat{a}\da \hat{a} + \frac{1}{2} \lambda_s \hat{a}\da \hat{a}\da \hat{a} \hat{a},
\end{align}
where:
\begin{align}
	\hbar \omega &:= \hbar \omega_0 + \frac{3}{4} m \omega^2_0 R^2,\\ \label{eq:lambdas}
	\lambda_s &:= \frac{g}{2 \sqrt{2} \pi^{3/2} R^3} \equiv \sqrt{\frac{2}{\pi}} \frac{a_s \hbar^2}{m R^3}.
\end{align}
Note that the MDDIs have  cancelled out due to the spherical symmetry of the BEC, leaving behind only the s-wave interactions \cite{PhysRevA.86.043625,tan2018effects}. This interaction term has the same form as the quantum gravitational interaction \eqref{eq:QGBEC}, and so we need to be able to distinguish between the electromagnetic and gravitational interactions in order to attribute  non-Gaussianity to only gravitational interactions. One way to achieve this is to use magnetic or optical Feshbach resonances, which are extraordinary processes particular to  cold atom and BEC experiments. Here  an external magnetic or optical field is used to resonantly couple a  molecular bound state  to a colliding atom pair, enabling the strength of the electromagnetic interactions to be  controlled \cite{RevModPhys.82.1225,PitaevskiiBook}.

Usually Feshbach resonances are used to increase the electromagnetic interaction strength between atoms in BECs. However,  they also allow  for the electromagnetic interaction to be in principle switched off, i.e.\ $\lambda_s = 0$, without affecting the strength of the gravitational interaction. This is achieved by applying a magnetic field of strength $B$ to the BEC, which results in the s-wave scattering length becoming a function of $B$ \cite{RevModPhys.82.1225,PitaevskiiBook}:
 \begin{align} \label{eq:Feshbach}
	a_s(B) = a^{bg}_s \Big[1 - \frac{\Delta}{B - B_0}\Big], 
\end{align}
where $a^{bg}_s$ is the background scattering length, $B_0$ denotes the resonance position and $\Delta$ is the resonance width. The s-wave interactions  can then be turned off  by setting $B = B_0 + \Delta$ \footnotetext[46]{{Rather than working with a spherical BEC where MDDIs cancel, we could alternatively use an harmonic trap to create a quasi-1D or quasi-2D geometry (extreme prolate or oblate spheroidal geometries) \cite{PitaevskiiBook}. In this case, instead of tuning the s-wave scattering length to zero, we can tune it to cancel the MDDIs such that we again have zero overall electromagnetic interactions \cite{PhysRevLett.101.190405}. We could also use  pure 2D or 1D Bose gases where the trapping geometry has one or two  dimensions with characteristic size much smaller than both two- and three-body length scales \cite{PitaevskiiBook}, leading to a drastic reduction in   collisional electromagnetic effects \cite{PitaevskiiBook}, but not gravitational effects \cite{Howl_2019}. In this case, there may be no phase transition to a BEC but there can still be macroscopic occupation of the ground state. Yet another option is to use $^{88}\mathrm{Sr}$ rather than  $^{133}\mathrm{Cs}$. This has zero total magnetic moment, meaning that it has no MDDIs and is insensitive to external magnetic fields. It also has a very small s-wave scattering length \cite{PhysRevLett.95.223002}, which is controllable through an optical Feshbach resonance \cite{FeshbachSr88,PhysRevLett.107.073202}.}}\cite{Note46}. For $^{133}\mathrm{Cs}$, this would be achieved when  $B= 17\unit{G}$ \cite{PhysRevA.70.032701,PhysRevA.79.013622}, leaving behind only the QG interactions \eqref{eq:QGBEC} which are unaffected by the applied magnetic field. With the electromagnetic interactions in the BEC turned off, non-Gaussianity can in principle be solely attributed to QG interactions in the BEC \footnotetext[47]{{Rather than  eliminating the noise from the electromagnetic interactions, we could use distinguishing features between the electromagnetic and gravitational interactions, such as how their strengths vary with the trapping potential and the magnitude  of an applied magnetic field.  For example, for a spherical trap, the effective strength of the electromagnetic interaction scales as $1/ R^3$, whereas gravity scales as $1/R$ - see \eqref{eq:lambdas} and \eqref{eq:size}. The different scaling is due to the fact that the electromagnetic interactions are already partially screened due to the atoms having overall neutral charge,  whereas the coupling of gravity is universal and so cannot be screened. With an applied magnetic field, it is even possible to change the s-wave interactions from being repulsive to attractive and vice-a-verse, which is clearly impossible to achieve with gravity.}}\cite{Note47}.

\section{Discussion}

\subsection{Role of Planck mass in proposed experiment}

We have argued that, as long as we are working in a situation where all other non-gravitational quantum interactions can be excluded, the production or change in  non-Gaussianity in the state of the quantum field of matter  would be sufficient evidence of QG, and have demonstrated how this could be used in a test that is based on just a single-well BEC. The size of the effect in the BEC experiment appears to be similar to that observed in the BMV proposal, see \eqref{eq:size} and \cite{BoseQGExp,BoseLocal,VedralQGExp,PhysRevD.98.046001}. This  illustrates how the experiment is related to the Planck mass since, using \eqref{eq:size}, we can write the SNR for one measurement in this case as \cite{RovelliQGExp,RovelliTime}:
\begin{align} \label{eq:tp}
\frac{M}{M_P} \frac{\delta \tau}{t_P}, 
\end{align}
where $M_P$ is the Planck mass, $t_P$ is the Planck time, and $\delta \tau := \sqrt{2 / \pi} G M t / (R c^2)$. This expression can also be derived by dividing the BEC into two halves, considering the gravitational interaction  of one with the other and the time dilation $\delta \tau$ induced in GR in the centre of each half. If we fix this SNR of one measurement, then \eqref{eq:tp} illustrates that as $M$ gets closer to $M_P$, it seems that we can probe more minute gravitational field intensities and thus further access its possible quantum properties.

\subsection{Direct classical gravity interactions can create entanglement between two quantum field systems but not non-Gaussianity} \label{sec:DirectInts}

A classical interaction can  create entanglement if this involves the respective quantum systems directly interacting with each other \cite{VedralQGExp,BoseQGExp,PhysRevD.98.046001,BoseLocal}. For example, consider two BECs that are in the two spatial arms of a  double-well potential. In the two-mode approximation, we can write the full quantum field of the atoms as $\hat{\Psi} (\bs{r}) = \psi_L (\bs{r}) \hat{a}_L + \psi_R (\bs{r}) \hat{a}_R$ where $\hat{a}_L$ and $\hat{a}_R$ respectively destroy an atom in the left  and right well, and $\psi_L$ and $\psi_R$ are the corresponding mode wavefunctions \cite{PitaevskiiBook,TwoModeReview}. In the case of CG, and taking the Newtonian approximation for simplicity, there will, in principle, be terms of the form $\lambda_{LR} \hat{a}\da_L \hat{a}_R + h.c.$, in the Hamiltonian, where $\lambda_{LR} := m \int d^3 \bs{r} \psi^{\ast}_L (\bs{r}) \psi_R (\bs{r}) \Phi [\varPsi](\bs{r},t)$. These are beam-splitting terms such that, if $\lambda_{LR}$ is non-zero due to, for example, the mode wavefunctions overlapping, and either BEC is in a non-classical state, then the terms  will induce entanglement between the BECs. There is an electromagnetic analogue of this effect where  a double-well trapping potential, which is approximated to be classical, causes or contributes to entanglement between the two wells. This entangling process is often referred to as ``quantum tunnelling'' in cold atom experiments \cite{PitaevskiiBook}.   However, since the entangling-inducing terms are quadratic, they will not induce non-Gaussianity, illustrating that, although a direct classical interaction with matter can create entanglement, it cannot create non-Gaussianity in the quantum field of matter. 

Note that here we are  working with  ``mode'' entanglement, i.e.\ entanglement between modes of a quantum field. If instead we attempted to use a  first-quantization picture and describe the full system using a many-body wave-function, then it is possible to argue that the initial state of the full system is  already entangled  and that Newtonian CG is not creating entanglement in this picture \cite{Lange416,Kunkel413,Fadel409}. This is because there is so-called ``particle'' entanglement before and after the effective CG beam splitter \cite{PhysRevLett.112.150501,QuantumLimitsOptics}. For example, the initial state could be $|\alpha\rangle_L |\xi \rangle_R$, with $|\alpha\rangle$ a coherent state and    $|\xi \rangle$ a squeezed  state, which, in a quasi first-quantization picture,  is particle entangled but not mode entangled \cite{QuantumLimitsOptics}. This occurs because, in the first quantization picture, a beam splitter does not couple the  left and right wells. However, in the relativistic CG limit we would also have, in principle, two-mode squeezing operations such as  $\hat{a}_L \hat{a}_R + \hat{a}\da_L \hat{a}\da_R$, that can result in a two-mode squeezed state, which is particle entangled \cite{QuantumLimitsOptics}. Therefore, in full generality and in either picture, CG can, in principle, create entanglement.  In contrast, in the first-quantization picture, it is possible for Newtonian CG to create non-Gaussian ``particle'' Wigner functions.  For example, the many-body wavefunction of our  single-well BEC experiment could start off Gaussian but become non-Gaussian under CG (see Appendix \ref{app:1stQuant} for more detail). However, in the more fundamental second quantization picture, the state of matter i.e., the state of the quantum field of matter, always remains Gaussian under CG, as  shown in Section \ref{sec:TheTheory}.


\subsection{Non-quantum interactions and continuous-time measurements}

Above we  defined a classical interaction as an interaction with an entity $\mathcal{G}$ that takes on real and well-defined values, such as the gravitational field of GR. We now consider  whether non-Gaussianity can also be  used to distinguish other, more general, non-quantum interactions  from their quantized counterparts. First we consider that $\mathcal{G}$ takes on complex values. This allows for the possibility that, most generally, the interaction can give rise to a Hamiltonian of the form \eqref{eq:Hquad} but where now the coupling constants $\bs{\lambda}_k$ and $\bs{\mu}_k$ are complex-valued. Although this, in general, leads to a non-Hermitian Hamiltonian, a Gaussian matter state  will continue to be of Gaussian form \cite{PhysRevA.83.060101,Graefe_2012,Graefe_2015},   and so non-Gaussianity can also distinguish this interaction from a quantum interaction\footnotetext[16]{{In general, the matter state will no longer be normalized, and by Gaussian ``form'' we mean a state that becomes Gaussian when it is normalized \cite{PhysRevA.83.060101,Graefe_2012,Graefe_2015}. Note that the physical density matrix can be defined as $|\varPsi\rangle \langle  \varPsi| / \langle \varPsi| \varPsi \rangle$, such that the final state would, in principle, be  automatically normalized. However, it is possible that $\langle \varPsi| \varPsi \rangle$ is ill-defined, and here we assume that  any such unphysical states that may result from the action of the non-Hermitian Hamiltonian are forbidden, perhaps due to an additional constraint on the interaction \cite{PhysRevA.83.060101,Graefe_2012,Graefe_2015}.}} \cite{Note16}.

Another possibility is that $\mathcal{G}$ could be a non-quantum but stochastic quantity. For example, a relativistic  theory of gravity coupled to matter has been proposed where the non-quantum gravitational field is stochastic \cite{2018arXiv181103116O}. It is found that gravity and matter interact through a  Gaussian completely-positive (CP)   channel, and so  non-Gaussianity should also rule out this non-quantum theory of gravity\footnotetext[17]{A Gaussian CP map is defined to be a CP map that takes any Gaussian state to another Gaussian state. The Clifford group \cite{PhysRevLett.88.097904}, is defined to be the group of Bogoliubov transformations, whereas the  Clifford semigroup \cite{PhysRevLett.89.207903},  is  the set of Gaussian CP maps. As with Bogoliubov transformations, the Clifford semigroup involves linear transformations of the quadratures, but the commutation relations do not have to be preserved.  In \cite{2018arXiv181103116O}, a Lindblad master equation with linear (matter) Lindblad operators is obtained for the interaction of matter with gravity, which is known to be a Gaussian CP map \cite{QuantifyingDecoherenceRef,GaussianStatesInContinuousQI}.} \cite{Note17}. More generally,  interacting a stochastic entity  $\mathcal{G}$ with a quantum field will still result in a Gaussian state of the quantum field  remaining Gaussian if we now broaden our definition of a Gaussian state to include states that are a statistical mixture of pure Gaussian states (the so-called Gaussian convex hull  \cite{ConvexHull,PhysRevA.90.013810}) \cite{Note18}. This is because a Gaussian state evolves to a state in the Gaussian convex hull if there is a combination of Gaussian operations and statistical randomization, i.e.\ stochasticity\footnotetext[18]{See Appendix \ref{app:StochInts} for more detail.} \cite{PhysRevA.87.062104}.

The preservation of this broader definition of Gaussianity also applies if the entity $\mathcal{G}$ is both stochastic and  complex-valued. However, in this case the norm will not, in general,  be  preserved, and so we have a mixture of unnormalized Gaussian states  \cite{PhysRevA.83.060101,Graefe_2012,Graefe_2015}. To ensure that the theory is norm-preserving, the physical state vector can be redefined as $|\varPsi \rangle/ ||\varPsi \rangle|$, which then allows for  a convex mixture of properly normalized Gaussian states. However, this, in general, results in a theory that is non-linear in the density matrix, leading to superluminal signalling \cite{Gisin1989}. Such an issue is also found in objective-collapse theories and, to rectify it,  a new higher-order process is applied to the evolution of the quantum system \cite{PhysRevD.96.104013,adler2004quantum}, which would here be associated with a quantum (self) interaction of matter i.e., a new force (See Appendix \ref{app:StochInts} for more detail and, in particular, Section \ref{app:PreservingNorm}). For example, if we represented matter by a real scalar field $\hat{\phi}$, the Hamiltonian of this new process  would involve terms that are quartic in $\hat{\phi}$. This new quantum process can, in general, induce non-Gaussianity. However, in the conventional case that the noise term of the objective-collapse theory has a Gaussian  profile and is anti-Hermitian (equivalent here to only the imaginary component of $\mathcal{G}$ being stochastic),  Gaussianity in the  matter field is still preserved. This is also  analogous to a continuous-time measurement being performed on matter by the stochastic entity $\mathcal{G}$, which could be a stochastic gravitational field, and the new quantum self-interaction\footnotetext[28]{Note that it is not possible for the  interaction with the stochastic entity $\mathcal{G}$ to perform continuous-time measurements of matter alone: a  higher-order process, as discussed in the main text, must also be introduced (see Appendix \ref{app:StochInts} and, e.g.\,\cite{PhysRevA.42.5086,CtsMeasurements} for more detail).}\footnotetext[21]{As illustrated in Appendix \ref{app:StochInts}, even when both the real and imaginary components of $\mathcal{G}$ are stochastic and obey a Gaussian probability distribution functional,  the state of matter will still asymptotically tend to a Gaussian state, despite the presence of the new quantum self-interaction. Of course, since objective-collapse theories are  introduced in an attempt to explain the so-called measurement problem of QT (see e.g.\ \cite{GRW2,DIOSI1988419,BELAVKIN1989355,CSL,Gisin_1992,Diosi_1988,PhysRevA.42.5086,CtsMeasurements,joos2013decoherence,PhysRevD.96.104013,adler2004quantum,DIOSI1985288,Kafri_2014} for gravitationally-inspired theories),  the state of matter must always tend to a classical-like state in these theories.

\hspace{0.3cm} When both the real and imaginary components are stochastic and there is a process related to  quantum self-interaction, then the resulting theory is closely related to  a continuous-time measurement being performed by the two interactions  but now with a feedback mechanism \cite{PhysRevA.42.5086}. Weak measurements with local feedback operations are also known to induce entanglement if the measurement is a joint  measurement \cite{PhysRevA.77.014305,PhysRevA.92.062321}.} \cite{PhysRevA.42.5086,Di_si_2015}\cite{Note28,Note21}.

\subsection{Alternative theories of gravity}

Einstein's GR can be formulated as an action theory with the action principle being used to derive the field equations. The action $S$ of GR can be decomposed into the Einstein-Hilbert action $S_{EH}$, that only contains gravitational degrees of freedom, and the matter action $S_M$, which tells us how matter and gravity interact:
\begin{align}
S = S_{EH} + S_M,
\end{align}
where:
\begin{align}
S_{EH} =  \frac{c^4}{16 \pi G} \int d^4 x \sqrt{g} R,
\end{align}
with $R$ the Ricci scalar; and for a real scalar matter field $\phi$:
\begin{align}
S_M = \frac{1}{2} \int \sqrt{g} \Big[ g^{\mu \nu} \partial_{\mu} \phi \partial_{\nu} \phi - (m^2 + \varepsilon R) \phi^2 \Big],
\end{align} 
with $\varepsilon$ a numerical factor. The matter actions for spin-1/2 and spin-1 fields are provided in Appendix \ref{app:GRAction}.

The argument that we have presented for non-Gaussianity being an indicator of QG only uses the matter action $S_M$ and says nothing of the purely gravitational action $S_{EH}$. It relies on the fact that $S_M$ must be, in the absence of all non-gravitational quantum interactions, quadratic in the matter fields such that a classical theory of gravity will preserve Gaussianity. If $S_M$ had terms that coupled gravitational degrees of freedom with  non-linear or non-quadratic functions of the matter fields then, in flat spacetime, such terms would still exist and this would result in a new non-gravitational interaction, which we have excluded.

Many theories of gravity have been suggested as alternatives to Einstein's GR \cite{CLIFTON20121}. These tend to consider alternative forms for the gravitational action $S_{EH}$. For example, in $f(R)$ theories of gravity \cite{fRGravity}, the $R$ in $S_{EH}$ is replaced with some function of the Ricci scalar $f(R)$. Using the argument above, as long as we can exclude all other relevant quantum interactions, non-Gaussianity can still be used as evidence of a quantum rather than classical version of these alternative theories of gravity.

\subsection{Non-Gaussianity in cosmology} \label{sec:CMB}

Non-Gaussianity is often considered in the context of cosmology. Here, studies  predominately concentrate on how the temperature fluctuations of the cosmic microwave background (CMB)  could follow a non-Gaussian distribution. So far measurements are consistent with a Gaussian distribution, but with the ever increasing precision of CMB measurements, it is possible that  non-Gaussianity could be detected in the near future, providing important insights into structure formation in our Universe.

Perhaps the most important mechanisms responsible for generating a non-Gaussian distribution of temperature fluctuations are inflationary mechanisms, which involve processes that occur at the surface of last scattering or before. These are often referred to as primary or primordial effects, and  can be further sub-categorized into quantum mechanical effects at or before horizon exit, and classical non-linear effects after horizon exit (see e.g\ \cite{InfaltionCourse}).  The most primitive physical mechanism for inflation assumes a single scalar field, the inflaton, that couples to gravity and obtains  a non-zero vacuum expectation value, leading to exponential expansion of space. The Hamiltonian for this model is similar to that with which we  used to illustrate that non-Gaussianity in the quantum state of  matter   can be used as evidence of QG (see \eqref{eq:Hphi} and \eqref{eq:LQGHs}). An important difference, however, is that the inflaton, unlike normal matter (Standard Model leptons and quarks),  has a non-quadratic potential $V(\hat{\phi})$ and thus self-interacts.

If the vacuum state is initially assumed then, at second order in quantum fluctuations of the inflaton and gravitational fields, a squeezed Gaussian state of curvature perturbations is created \footnotetext[37]{See e.g.\ \cite{Guerreiro_2020,WilczekSqueezeGWs} for recent proposals on observing squeezed gravitational waves with gravitational wave detectors.}\cite{Note37}, which is the leading order effect and can explain the Gaussian nature of the CMB that has been observed so far (see e.g\ \cite{InfaltionCourse,martin2005inflationary}). At the next order, non-Gaussian effects occur due to the coupling between QG and the inflaton as well as self-interactions of the inflaton \footnotetext[36]{This is in fact a gauge-dependent statement since we can always choose the unitary gauge where quantum fluctuations of the inflaton vanish, and non-Gaussianity of curvature perturbations is then due entirely to non-linear QG effects \cite{InfaltionCourse}.}\cite{Note36}. The former effect is analogous to the interaction we have considered between QG and matter that generates non-Gaussianity in the quantum state of matter. The latter effect means that, in principle, it would be possible for purely classical gravitational effects to enhance  non-Gaussianity in this inflaton model since gravity will couple to the non-Gaussian-generating self-interaction of the inflaton. This further illustrates that only if we are able to work in a situation where all other quantum mechanical interactions can be ignored can we use non-Gaussianity as evidence of a quantum theory of gravity. In this case, we are not neglecting the quantum self-interaction of the inflaton. Furthermore, unless a proper measure of non-Gaussianity is used, then even if we can neglect the inflaton self-interactions, CG effects can  give the \textit{appearance} of enhancing any already present non-Gaussianity in the inflaton. This is because the Hamiltonian \eqref{eq:Hphi} can lead to an increase or decrease in higher-order cumulants, such as $\kappa_4$, if the scalar field is already in a non-Gaussian state. However, when a proper measure of non-Gaussianity is used, such as the SNR defined in \eqref{eq:SNR} or that based on quantum relative entropy \cite{PhysRevA.78.060303}, these measure do not change, and so  non-Gaussianity is not really increasing or deceasing.

As mentioned above, non-Gaussianity in quantum curvature perturbations can also occur after horizon exit. In this case, a non-linear classical evolution can result in a non-linear relationship between the quantum curvature perturbations and the inflaton field, which results in  non-Gaussian statistics for the  curvature perturbations even if the inflaton is in a Gaussian state (see e.g.\ \cite{InfaltionCourse}). Note that here the non-Gaussianity is in the quantum curvature perturbations not in the inflaton field, which, if we neglect its self-interactions, is analogous to the matter field $\phi$ we have used in Section \ref{sec:TheTheory} to illustrate our argument that CG cannot create non-Gaussianity in the quantum field state of matter.

The reason that non-Gaussianity in the quantum state of curvature perturbations is related to non-Gaussianity in the temperature fluctuations of the CMB is due to the Sachs-Wolfe (SW) effect \cite{SachsWolfe}. This is a classically treated effect where the curvature fluctuations result in red-shifts to the radiation such that there is a direct map between correlation functions of the curvature perturbations and correlation functions of temperature fluctuations in the CMB. Any non-Gaussianity  in the CMB due to a single-field model of inflation is expected to be very small and outside the realms of near-future observations of the CMB. Instead, more complex models are needed for observable levels of non-Gaussianity, such as multi-field inflation \cite{Battefeld_2007}. Detection of non-Gaussianity would, therefore, potentially provide crucial knowledge of the mechanisms responsible for structure formation.

However, there are many other mechanisms responsible for creating a non-Gaussian distribution of temperature fluctuations in the CMB besides the inflationary ones. These include so-called secondary effects, which involve  processes that occur between the last scattering surface and the observer (see e.g\ \cite{yadav2010primordial,Bartolo_2006}). Secondary effects can be broadly divided into  scattering of the CMB radiation with hot gas,  and classical effects mediated by gravity subsequent to the last scattering surface, such as the time-integrated SW effect \cite{SachsWolfe} and gravitational lensing. Other effects that can be responsible for non-Gaussian temperature fluctuations include non-linear effects at recombination (see e.g.\ \cite{yadav2010primordial,Bartolo_2006}). Although detection of these non-primordial effects would provide important information for  distinguishing structure formation scenarios, they are often regarded as noise of the primary inflationary effects.

It is thought that measurements of the CMB could provide evidence for a quantum theory of gravity. For example, QG predicts that there will be tensor modes due to quantum fluctuations of the gravitational field during inflation. Dimensional arguments involving Planck's constant can then potentiality be used for evidence of QG if such modes are observed \cite{PhysRevD.89.047501}. Other possibilities have also been suggested, such as using measurements of the scalar modes of the temperature fluctuations to try and access quantum measures of the primordial curvature perturbations, such as violation of Bell inequalities \cite{PhysRevD.96.063501} or quantum discord \cite{QDiscordCMB}. The issue here, however, is that we only have access to the field modes, not their momentum-conjugate, which is part of the `decaying mode', and we can only measure one instance of the CMB at a time. There  then does not seem to be enough information to rule out classical curvature perturbations \cite{MaldacenaCMB,QDiscordCMB,PhysRevD.96.063501}, and instead rather convoluted inflationary models need to be assumed for evidence of QG \cite{MaldacenaCMB}.

If we were to brute forcefully apply our non-Gaussianity argument for evidence of QG to cosmology, then this would require measuring the  quantum state of the CMB radiation and somehow being able to distinguish gravitational interactions with the CMB from non-gravitational ones, such as secondary effects due to scattering with electrons. In this case, assuming that the initial quantum state of the CMB is Gaussian, any sign of non-Gaussianity in its quantum state due to gravitational interactions would be evidence of a quantum theory of gravity. Unfortunately, measuring the temperature fluctuations of the CMB radiation does not, in general, provide information on the quantum state of the radiation. For example, if we were to consider an ideal gas of radiation in a container in a static curved spacetime, then  at thermal equilibrium the radiation has  Bose-Einstein statistics \cite{StatDist,chavanis2020statistical} and  the quantum state of the radiation is a Gaussian thermal state (see e.g.\ \cite{GQI}). However, due to the form of the spacetime metric and the Ehrenfest–Tolman effect \cite{PhysRev.35.904,PhysRev.36.1791}, the radiation can have temperature fluctuations in space that obey a non-Gaussian distribution. Therefore, non-Gaussian fluctuations of the temperature of the CMB do not necessarily mean that the quantum state of the CMB itself is  non-Gaussian. Any fundamentally classical gravitational effects that generate non-Gaussianity in the temperature fluctuations of the CMB will not change the non-Gaussianity in the quantum state of the radiation. This applies in the perturbative as well as non-perturbative regime of the gravity since our argument is just  based on the fact that the  CG Hamiltonian is  quadratic in the quantized matter fields, which is the case in both  the perturbative and non-perturbative regimes of gravity (this also means that the CG Hamiltonian can be solved in the absence of  standard quantum perturbation theory - see Appendix \ref{app:CG}).  Instead, quantum gravitational mechanisms would need to be considered, such as a quantum version of the SW effect, for changing the non-Gaussianity of the quantum state of the CMB radiation due to gravity.

Measurements of temperature fluctuations of the CMB are thought to provide information on the quantum state of the primordial  gravitational field through the classical SW effect. In contrast, if we were able to measure the non-Gaussianity of the quantum state of the CMB radiation and robustly claim that this is due to gravitational interactions, then this would provide evidence of QG through an indirect means of analysing the quantum state of radiation, with no knowledge of the state of the primordial gravitational field required.

\subsection{Applicability}

We have argued that, when representing matter with  a quantum field, non-Gaussianity in its quantum state  can be used as an indirect signature of a quantum rather than classical theory of gravity. In certain theories of QG, such as loop QG, group field theory \cite{Oriti2007,freidel2005group} and asymptotically safe QG \cite{weinberg1978critical,weinberg1979ultraviolet},  matter is fundamentally described by quantum fields. However, in other QG theories, such as string theory, this representation of matter is a limiting low-energy description of that used in the fundamental theory, and the low-energy description is referred to as `effective field theory' \cite{EFTBurgess,EFTGeorgi}. In this case, if we wanted to use a notion of non-Gaussianity that is applicable to the representation used in the full theory, our concept of non-Gaussianity in matter would have to be updated, or it may only be applicable to the low-energy effective field theory description. Our argument stems from the fact that the Hamiltonian or action of gravity in the quantum field regime of matter has matter-gravity terms that are  only quadratic in the  quantum matter  operators. There is a connection here with string theory, see the Polyakov action in curved space \cite{superstring1}, suggesting that our notion of non-Gaussianity in matter could also be generalized to  strings. However, in foreseeable table-top tests of QG, it is very unlikely that anything beyond the (potentially effective) quantum field theory description QG will be probed.   


\section{Summary}

We have shown that, if we want to create non-Gaussianity in the quantum field state of matter with purely gravitational interactions, then this would not be   possible with a classical theory of gravity, but can be achieved with a quantum theory of gravity.  On the theoretical side, this opens up a new   connection between QG and QIS. Perhaps the most important property of this new connection is that, in contrast to other quantum information resources such as entanglement, although QG can create non-Gaussianity, CG can never create non-Gaussianity in the quantum field of matter (as long as, as with entanglement, all other quantum interactions can  be neglected). For example, while a classically expanding spacetime metric can create entanglement in the quantum field of matter \cite{PhysRevLett.21.562,PhysRev.183.1057,PhysRevD.3.346,BirrelandDavies,Entanglementinthesecondquantizationformalism,BALL2006550,PhysRevD.82.045030,PhysRevD.89.024022}, it cannot create non-Gaussianity. This also suggests that, whereas entanglement is not invariant to changes in classical reference frames \cite{RevModPhys.76.93}, non-Gaussianity and non-classicality are.   

Non-Gaussianity is a very important resource in QIS. For instance, it is necessary for universal quantum computation \cite{PhysRevLett.82.1784,PhysRevA.68.042319}. However,  it is not sufficient for universal speed-up over classical computation. For this, we need negative Wigner function states \cite{PhysRevLett.109.230503,Veitch_2013}, and in the case of mixed states, it is possible for a non-Gaussian state to have a positive Wigner function. Given that it is negative Wigner function states that are generically associated with non-classicality, it is interesting from a fundamental point-of-view that it appears to be non-Gaussianity that is a universal indicator of QG rather than negative Wigner function states. Perhaps non-Gaussianity in matter, especially  its broader definition as states outside the Gaussian convex hull, is connected with a fundamental property of quantum gravitational degrees of freedom, such as non-commutating variables or quantum contextuality, which will be studied in future work.
 
Approaching QG from a quantum information perspective has attracted much theoretical interest recently, especially in regards to uncovering the role that quantum correlations, such as entanglement, may play in the fundamentals of QG. Conventionally QG has been considered in the context of discrete-variable quantum information, such as qubits,  whereas, here we have concentrated on describing QG using continuous-variable quantum information, and resources particular to CVQIS. Just as CVQIS has been extremely effective in connecting quantum information and quantum field theory, we find that it is also very powerful in connecting quantum information  and QG.  Our findings, however, could also potentially be extendable to describing QG with  discrete-variable quantum information since the Wigner function can also  be defined for discrete systems \cite{WOOTTERS19871,Vourdas_2004,PhysRevA.65.062309,PhysRevA.70.062101}.


As well as providing a theoretical link between QG and QIS, we have also shown how non-Gaussianity can be used to implement new types of experimental tests of QG. In particular,  non-Gaussianity allows for tests based on just a single rather than multi-partite quantum system, and it also provides a particularly reliable signature of QG since it cannot be created by indirect or direct CG interactions (as long as all other interactions can be neglected). This is in contrast to previous tests based on entanglement where a multi-partite quantum system is necessary and where entanglement is only an indicator of QG in certain contexts, allowing, in principle, for certain loopholes in which CG could be creating the expected QG signal, such as non-local effects or direct CG interactions that occur due to overlapping  mode wavefunctions  (see Section \ref{sec:DirectInts}).

We have provided a  proposal for a table-top test of QG that uses our non-Gaussianity witness and that should be achievable in the near future. This proposal uses a $4 \times 10^9$ BEC in just a single-well potential, with $10^9$ atomic BECs  having already been created in single wells \cite{HydrogenBEC}.  The most promising proposal so far for a table-top test of QG is considered to be the BMV proposal, which, in contrast to our quantum gas experiment, uses an opto-mechanical setup and  entanglement as a witness of QG. In our proposal and the BMV proposal, the QG signal scales quadratically with the mass of the system and linearly with the interaction time.  
 The signal in both proposals is greatest when the initial state is a highly non-classical state: here we have used a squeezed state, whereas the BMV effectively uses a N00N state, which is considered to be the most challenging quantum state to create. So far, neither N00N states nor squeezed states in the quantum regime  have  been created in nano/micro-particle experiments, whereas squeezed states in the quantum regime have been created in BECs \cite{Linnemann_2017}, further facilitating the implementation of our proposal. However, macroscopic quantum squeezed states have yet to be generated. We have investigated how these could be achieved in the near term, but, just as with other proposed table-top tests of QG that use QIS \cite{Ulbricht20015,Carlesso_2019,BoseQGExp,BoseLocal,VedralQGExp,PhysRevD.98.046001,PhysRevLett.119.120402,krisnanda2020observable}, creating such macroscopic states is  an experimental challenge in realizing the experiment. Another option  would  be to  use a classical coherent initial state but increase the mass of the system and the number of repetitions of the experiment by an order or two of magnitude, which will be considered in future work.

As with other recently proposed QIS table-top tests of QG \cite{kafri2013noise,Ulbricht20015,Carlesso_2019,BoseQGExp,BoseLocal,VedralQGExp,PhysRevD.98.046001,PhysRevLett.119.120402,Kafri_2014,krisnanda2020observable}, we  need to ensure that the electromagnetic interactions can be suppressed or distinguished from the gravitational interactions. In the BMV proposal, for example, this means increasing the separation between the microspheres, which also lowers the gravitational signal, as well as ensuring that there is no excess charge on either microsphere. An advantage in using quantum gases is that the electromagnetic interactions are  manipulable using external magnetic or optical fields \cite{PitaevskiiBook}, which allows the noise from electromagnetic interactions to be suppressed  without affecting the strength of the gravitational interactions. We have considered a $^{133}\mathrm{Cs}$ BEC since this has the broadest and strongest Feshbach resonance, allowing for, in principle, sufficiently low levels of electromagnetic noise in the experiment. 

BECs and cold atoms have  been found to be very effective in tests of classical gravity, and experiments using these are now becoming the state-of-the-art for many types of classical gravity measurements \cite{bongs2019taking}. Their effectiveness can be attributed to the extraordinary degree of control that BECs and cold atoms provide in exploring  the fundamental behaviour of quantum matter in various settings. In particular, Feshbach resonances enable  control of the electromagnetic  interactions between  atoms, providing a key tool that has led to several scientific breakthroughs \cite{RevModPhys.82.1225,PitaevskiiBook}. Given their great success in classical gravity measurements and the degree of control offered by these systems, it is perhaps not surprising that we  find that BECs could also be very useful for measuring \emph{quantum} gravitational effects. 




\begin{acknowledgments}


We thank Chiara Marletto, Andrea Di Biaggio and participants of the  Quantum Information Structure of Spacetime (QISS) HKU 2020 workshop and  Physical  Institute  for  Theoretical  Hierarchy (PITH) for stimulating and insightful  discussions. R.H.\ acknowledges ERC StG GQCOP (Grant Agreement No.\ 637352), as well as Gerardo Adesso and Giulio Chiribella for their great  support. R.H. also acknowledges Jan Robbe for generously allowing the use of his artwork ``Space and Particles 8'' for a creative visualization of the proposed BEC experiment. V.V., M.C.\ and C.R.\ acknowledge the support of the ID 61466 grant from the John Templeton Foundation, as part of the \href{http://www.qiss.fr}{QISS project}. The opinions expressed in this publication are those of the authors and do not necessarily reflect the views of  the John Templeton Foundation.  D.N.\ acknowledges the support of ``Agence Nationale pour la Recherch'' (grant EOSBECMR \# ANR-18-CE91-0003-01). 

\end{acknowledgments}

\newpage

\onecolumngrid

\appendix

\section{Non-Gaussianity in quantum gravity} \label{app:GRAction}

The way in which matter and gravity interact in GR is described by the matter action $S$, which can be derived from the specific Lagrangian density $\mathcal{L}(x)$ for the matter field:
\begin{align}
S = \int d^4 x \mathcal{L} (x) .
\end{align} 
For example, neglecting all other interactions (which automatically includes any self-interactions), then in the metric or tetrad formulations of GR, the respective Lagrangian densities for a real scalar $\phi$, spin-1/2 $\psi$, and spin-1  field $A_{\mu}$, are \cite{rovelli_2004,BirrelandDavies}:
\begin{align}\label{eq:Lscalar}
\mathcal{L}_{\phi} &= \frac{1}{2} \sqrt{g} [g^{\mu \nu} \partial_{\mu } \phi  \partial_{\nu} \phi - (m^2 + \varepsilon R) \phi^2]\\
&\equiv \frac{1}{2} e [\eta^{\alpha \beta} e^{\mu}_{\alpha} \partial_{\mu } \phi e^{\nu}_{\beta} \partial_{\nu} \phi - (m^2 + \varepsilon  R) \phi^2],\\
\mathcal{L}_{\psi}  &= \sqrt{g} \left(\frac{1}{2} i [ \overline{\psi} \gamma^{\mu} \nabla_{\mu} \psi - (\nabla_{\mu} \overline{\psi}) \gamma^{\mu} \psi] - m \overline{\psi} \psi \right),\\ \label{eq:Lvector}
\mathcal{L}_{A}  &= -\frac{1}{4} \sqrt{g} g^{\mu \nu} g^{\nu \sigma} F_{\mu \rho} F_{\nu \sigma},
\end{align}
where $e^{\mu}_{\alpha} (x)$ are tetrads, the `matrix square root' of the metric tensor: $g^{\mu \nu} (x) =: e^{\mu}_{\alpha} (x) e^{\nu}_{\beta} (x) \eta^{\alpha \beta}$, with $\mu$ labelling the general spacetime coordinate, $\alpha$ the local Lorentz spacetime,  and $\eta^{\alpha \beta}$ is the Lorentz metric. Furthermore, $F_{\mu \nu} := \partial_{\nu} \mathcal{A}_{\mu} - \partial_{\mu} \mathcal{A}_{\nu}$
is the electromagnetic tensor; $\mathcal{A}_{\mu}$ is the electromagnetic four-potential; $R$ is the Ricci scalar; $\nabla_{\mu}$ is the covariant derivative;  $\gamma^{\mu} := e^{\mu}_{\alpha} \gamma^{\alpha}$ are the curved space counterparts of the gamma (Dirac) $\gamma$ matrices,  which satisfy $\{ \gamma^{\mu}, \gamma^{\nu}\} = 2 g^{\mu \nu}$; $\varepsilon$ is  a numerical factor which we set to zero for the rest of this Appendix for simplicity;  and the chosen metric signature is $(-,+,+,+)$. Note that for a complex rather than real scalar field, we just replace terms with two copies of $\phi$ by one copy of $\phi^{\ast}$  and  $\phi$, e.g.\ $\partial_{\mu} \phi \partial_{\nu} \phi$ becomes $\partial_{\mu} \phi^{\ast} \partial_{\nu} \phi$.

We can also write corresponding Hamiltonian (constraint) densities for the above Lagrangian densities \cite{Thiemann_1998,rovelli_2004,thiemann2008modern}:
\begin{align} \label{eq:LQGHs}
\mathcal{H}_{\phi} &= \frac{1}{2}\left(\frac{\pi^2}{\sqrt{\mathfrak{g}}}+\sqrt{\mathfrak{g}} \mathfrak{g}^{ab} \partial_{a} \phi \partial_b \phi + \sqrt{\mathfrak{g}} m^2 \phi^2 \right),\\
\mathcal{H}_{\psi}&= \frac{1}{2 \sqrt{\mathfrak{g}}} E_{j}^{a} \left[ i \zeta \tau^j  \mathcal{D}_{a} \xi + \mathcal{D}_{a}(\zeta \tau^j \xi) + \frac{1}{2} i K^j_a \sigma \xi + c.c.\right],\\ \label{eq:LQGHsEnd}
\mathcal{H}_{A}&=\frac{1}{2 \sqrt{\mathfrak{g}}} \mathfrak{g}_{a b} \left[\mathcal{E}^{a} \mathcal{E}^{b}+\mathcal{B}^{a} \mathcal{B}^{b}\right].
\end{align}
Here spacetime  has been split into spatial slices and a time axis $M = \mathbb{R} \times \sigma$. Taking $n^{\mu}$ to be the normal vector field of the time slices $\sigma$, the tetrad can be written as $e^{\mu}_{\alpha} = \mathfrak{e}^{\mu}_{\alpha} - n^{\mu} n_{\alpha}$, with $\eta^{\alpha \beta} n_{\alpha} n_{\beta} = - 1$ an internal unit timelike vector (which we may choose to be $n_{\alpha} = - \delta_{\alpha, 0}$), so that $\mathfrak{e}^{\mu}_{\alpha}$ is a triad, where $\mathfrak{e}^{\mu}_{\alpha} = (0,\mathfrak{e}^{\mu}_i)$ and we further define $E^{a}_i = |\det \mathfrak{e}^{a}_i | \mathfrak{e}^{a}_i$ with $i,a = 1,2,3$. The conjugate momenta to the densitized triad $E^{a}_i$ is the chiral spin connection $A^i_{a} := \Gamma^i_{a} + K^i_{a}$, where  $\Gamma^i_{a} = \Gamma_{a jk} \epsilon^{jki}$ and  $K^i_{a} = K_{a b} \mathfrak{e}^{b i}$,  with  $\Gamma_{a jk}$  the spin-connection and $K_{a b}$  the extrinsic curvature. In \eqref{eq:LQGHs}-\eqref{eq:LQGHsEnd},   $\mathfrak{g}$  is then  the determinant of the induced spatial metric $ \mathfrak{g}^{ab} \equiv  \mathfrak{e}^a_i \mathfrak{e}^b_j \delta^{ij}$ on the spatial slices; $\pi := \sqrt{\mathfrak{g}} n^{\mu} \partial_{\mu} \phi$ is the momentum conjugate to $\phi$; $\mathcal{E}^a := \sqrt{\mathfrak{g}} \mathfrak{g}^{ab} n^{\mu} F_{\mu b}$ is the electric field; $\mathcal{B}^a := \epsilon^{abc} F_{bc}$ is the magnetic field; $\tau_i$ are the generators of the Lie algebra su(2) with the convention [$\tau_i, \tau_j] = \epsilon_{ijk} \tau_k$; $\xi = \sqrt{\mathfrak{g}} \psi$, with $\psi$ a Grassman-valued fermion field; $\zeta$ is the momentum conjugate to $\xi$; and $\mathcal{D}_a \xi := (\partial_a + \tau_j  A^j_a ) \xi$. For simplicity, we have also assumed that the scalar and fermionic fields are singlets under any internal group symmetry.

Since we have neglected all other interactions, the above Lagrangian and Hamiltonian densities are all necessary quadratic in matter fields as they then only consist of kinetic and mass terms.  This quadratic scaling of course applies to any  spin field not  just those considered above \cite{BELINFANTE1940305,PhysRev.94.1472,PenroseAnySpin,DowkerDowker1966a,Dowker_1966,DOWKER1972577,Grensing_1977,CHRISTENSEN1978571,Birrell_1979}. Therefore, if we quantize the matter fields but leave the gravitational degrees of freedom classical, we have a theory that preserves Gaussianity. However, if gravity obeys a quantum theory, then there must be some quantum operator associated with it, and, therefore, we must have  a theory that has interactions involving three or more quantum operators and that thus  induces non-Gaussianity. In the next two sections we also illustrate this argument  in the weak-field  and non-relativistic  limits of gravity. 

\subsection{Weak-field limit} \label{app:WeakField}

In the weak-field limit of gravity, we write $g_{\mu \nu} = \eta_{\mu \nu} +  h_{\mu \nu}$, where $h_{\mu \nu}$ is a perturbation around a space-time background with metric $\eta_{\mu \nu}$. In this case, the GR matter-gravity interaction Hamiltonian can be written as \cite{maggiore2008gravitational}:
\begin{align} \label{eq:HintRel}
H_{int} = -\frac{1}{2} \int d^3 \bs{r} \, T^{\mu \nu} h_{\mu \nu},
\end{align}
where
\begin{align} \label{eq:Boxh}
\Box h_{\mu \nu} = \frac{ 16 \pi G}{c^4} \Big(\frac{1}{2} \eta_{\mu \nu} \eta^{\sigma \rho} T_{\sigma \rho} -  T_{\mu \nu}\Big),
\end{align}
with $\Box$  the d'Alembert operator,   $T^{\mu \nu}$  the stress-energy tensor for matter, and we have chosen the Lorentz gauge. The stress-energy tensor for a field of arbitrary spin in curved spacetime can be obtained by variation of the action with respect to the metric \cite{BirrelandDavies}:
\begin{align}
T_{\mu \nu} (x) = \frac{2}{\sqrt{-g}} \frac{\delta S}{\delta g^{\mu \nu} (x)} \equiv\frac{ e_{\alpha \mu} (x)}{e(x)} \frac{\delta S}{\delta e_{\alpha}^{\mu} (x)}.
\end{align}
For example, when neglecting all other interactions, for a real scalar, spin-1/2 and spin-1 field, the curved space stress-energy tensors  are (before taking a weak-field limit) \cite{BirrelandDavies}:
\begin{align}
T^{\phi}_{\mu \nu} &= (1-2\varepsilon) \partial_{\mu} \phi \partial_{\nu} \phi + (2 \varepsilon -\frac{1}{2}) g_{\mu \nu} g^{\rho \sigma} \partial_{\rho} \phi \partial_{\sigma} \phi - 2 \varepsilon (\nabla_{\mu} \partial_{\nu} \phi) \phi + \frac{1}{2} \varepsilon g_{\mu \nu} \phi \Box \phi \\&~~~- \varepsilon \Big[R_{\mu \nu} - \frac{1}{2} R g_{\mu \nu} (1 - 3 \varepsilon)\Big] \phi^2 + \frac{1}{2} \Big[1- 3 \varepsilon\Big] m^2 g_{\mu \nu} \phi^2,\\
T^{\psi}_{\mu \nu} &= \frac{1}{2} i [\overline{\psi} \gamma_{(\mu} \nabla_{\nu)} \psi - [\nabla_{(\mu} \overline{\psi}] \gamma_{\nu)} \psi],\\
T^{A}_{\mu \nu} &= \frac{1}{4} g_{\mu \nu} F^{\rho \sigma} F_{\rho \sigma} - F^{\rho}_{\mu} F_{\rho \nu},
\end{align}
where we have ignored any gauge fixing or ghost terms in  $T^{A}_{\mu \nu}$ \cite{BirrelandDavies}.  Since we have neglected all other interactions, all stress-energy tensors  are necessarily just quadratic in matter fields.

In a QG theory we add a hat to both $T_{\mu \nu}$ and $h_{\mu \nu}$. This then results in an interaction Hamiltonian that is cubic in field operators. For example, for a complex scalar field we have terms of the form $\hat{\phi}\da \hat{\phi} \hat{h}_{\mu \nu}$, where we have suppressed any derivatives. On the other hand, for a CG theory, the interaction Hamiltonian contains terms only quadratic in quantum field operators. For example, in the semi-classical theory of gravity \cite{moller1962theories,ROSENFELD1963353}, with  complex scalar matter fields, we have terms of the form $\hat{\phi}\da \hat{\phi} h_{\mu \nu}$, where $h_{\mu \nu}$ is  given by the expectation value of the right-hand side of \eqref{eq:Boxh}. Therefore, this weak-field limit of CG cannot produce or  change  non-Gaussianity in the field state of matter, whereas QG can, as expected from the general discussion of GR and QG in the previous section.

\subsection{Newtonian limit} \label{app:NewtLimitQG}

We now consider a Newtonian theory of gravity with matter quantized. This can be obtained by starting from Newton's theory and quantizing matter  or from taking the non-relativistic limit of the  weak-field theory. For the latter, we consider  a \emph{closed system} and only the components $T_{00}$ and $h_{00}$. This results in Poisson's equation \cite{poisson_will_2014}:
\begin{align} \label{eq:Poisson}
\nabla^2 \Phi (\bs{r}) &= 4 \pi G \rho(\bs{r}) \\ \label{eq:PhiNG}
\implies \Phi (\bs{r}) &= - G \int d^3 \bs{r'} \frac{\rho(\bs{r'})}{|\bs{r} - \bs{r'}|},
\end{align}
and the Newtonian interaction Hamiltonian:
\begin{align} 
H_{int} = \frac{1}{2} \int d^3 \bs{r} \rho (\bs{r}) \Phi (\bs{r}),
\end{align}
where $\Phi := - c^2 h_{00} / 2$ is the Newtonian potential, and $\rho := T_{00} / c^2$ is the matter density. Irrespective of the spin of the field, $\rho$ again contains two copies of the matter field, e.g.,  for a single non-relativistic scalar matter field $\Psi$, $\rho = m \Psi^{\ast} \Psi $. The interaction Hamiltonians for quantum and classical Newtonian gravity (with quantized scalar matter fields) are then \eqref{eq:HintQGNewt}-\eqref{eq:HintCGNewt}:
\begin{align} 
\hat{H}_{QG}^{int} &= \frac{1}{2} m \int d^3 \bs{r} :\hat{\Psi}\da (\bs{r}) \hat{\Psi} (\bs{r}) \hat{\Phi} (\bs{r}):\\\nonumber
&= - \frac{1}{2} G m^2 \int d^3 \bs{r'} d^3 \bs{r} \frac{\hat{\Psi}\da (\bs{r'}) \hat{\Psi}\da (\bs{r})\hat{\Psi} (\bs{r'}) \hat{\Psi} (\bs{r})}{|\bs{r} - \bs{r'}|},\\ 
\hat{H}_{CG}^{int} &= m \int d^3 \bs{r} \hat{\Psi}\da (\bs{r})  \hat{\Psi}(\bs{r}) \Phi[\varPsi] (t,\bs{r}),
\end{align} 
where $::$ refers to normal ordering, and we have made explicit that $\Phi$ may depend on the quantum state of matter $\varPsi$ in a CG theory, which can result in single-particle gravitational self-interaction, for which we have dropped a factor of $1/2$. For example, for the Schr\"{o}dinger-Newton equations (the non-relativistic limit of semi-classical gravity), $\Phi$ is given by the expectation value of the right-hand side of the quantized version of \eqref{eq:PhiNG}. Expanding the non-relativistic field in annihilation operators, $\hat{\Psi}(\bs{r}) = \sum_k \psi_k (\bs{r})  \hat{a}_k$, we again find CG is  only quadratic in quantum operators and so cannot change the degree of quantum non-Gaussianity in the state of matter, whereas QG can.

\subsubsection{First quantization} \label{app:1stQuant}

The interaction Hamiltonian of classical Newtonian gravity is given by \eqref{eq:HintNG}. The Hamiltonian of QG and CG in the Newtonian limit can then be derived by quantizing the matter density $\rho(\bs{r})$ and, in the QG case, the gravitational potential $\Phi(\bs{r})$. In the previous section we took matter to obey a non-relativistic quantum field $\hat{\Psi}$, such that $\hat{\rho} = m \hat{\Psi}\da \hat{\Psi}$, assuming a single type of matter. Since $\hat{\Psi}$ is linear in  annihilation operators, and so also in  quadratures, the interaction Hamiltonian for CG is at most quadratic, such that an initial Gaussian state of the matter field will always remain Gaussian. 
However, in the case that we always have definite particle number, which can only be possible in the Newtonian approximation of the respective theories not the full relativistic theories, we could also view QG and CG in a first-quantized form\footnotetext[35]{Note that even in the Newtonian approximation it is still possible for a quantum system to have indefinite particle number, i.e.\ to not be an eigenstate of the number operator for the system. For example, the system could be an open quantum system.} \cite{Note35}. In this case, assuming a single type of particle, we may quantize $\rho(\bs{r})$ through:
\begin{align}
\hat{\rho}(\bs{r}) = m \sum_{i=1}^N \delta^{(3)} (\bs{r}-\hat{\bs{r}}_i),
\end{align}
where $N$ is the total number of particles in the matter system. The respective QG and CG Hamiltonians would  then be:
\begin{align}
\hat{H}^{int}_{QG} &= \frac{1}{2} m \sum^N_{i=1} \hat{\Phi} (\hat{\bs{r}}_i),\\ \label{eq:CG1stQuant}
\hat{H}^{int}_{CG} &= m \sum^N_{i=1}\Phi[\varPsi] (\hat{\bs{r}}_i).
\end{align}
Since $\Phi(\bs{r})$ does not need to be a quadratic function of $\bs{r}$, it is possible here for CG to create non-Gaussianity in the first-quantization picture. For example, in the Shr\"{o}dinger-Newton equations, where $\Phi(\bs{r}) = \langle \hat{\Phi}(\bs{r}) \rangle$ with $\hat{\Phi}(\bs{r})$ obeying Poisson's equation \eqref{eq:Poisson}, the many-body wavefunction of $N$ massive particles would evolve as \cite{Bahrami_2014}:
\begin{align}\nonumber
i \hbar \partial_t \psi_N(t; \bs{r}_1, \ldots, \bs{r}_N) = \Big(&- \frac{\hbar^2}{2m} \sum^N_{i = 1}  \nabla^2_i + V(\bs{r}_1, \ldots, \bs{r}_N)\\ &- G m^2 \sum^N_{i,j = 1} \int d^3 \bs{r}'_1 \cdots d^3 \bs{r}'_N \frac{|\psi_N (t; \bs{r}'_1, \ldots, \bs{r}'_N)|^2}{| \bs{r}_i - \bs{r}'_j|} \Big) \psi_N(t; \bs{r}_1, \ldots, \bs{r}_N),
\end{align}
where $V$ is a trapping potential. Although a Gaussian approximation is expected to be very good for table-top experiments \cite{PhysRevA.77.023619,grossardt2016effects}, the evolution of $\psi_N$ (and  its corresponding Wigner function) can, in principle, be non-Gaussian. Therefore, in the BEC experiment proposed in the main text, although the state of the BEC in the second-quantization picture must stay Gaussian under CG, its many-body wave-function need not.  Note also that, just as particles tend to automatically get ``entangled'' in the first-quantization picture when we have identical particles, an identical particle system also tends to become automatically non-Gaussian. That is, if we have two identical particles in positions $\bs{r}_1$ and $\bs{r}_2$ and two different states $a$ and $b$, then the many-body wavefunction is $\psi_N = [\phi_a(\bs{r}_1) \phi_b(\bs{r}_2) \pm \phi_a(\bs{r}_2) \phi_b(\bs{r}_1)] /\sqrt{2}$, depending on whether the particles are bosons or fermions. The system looks entangled just because of the  exchange symmetry of the identical particles (it is so-called ``particle'' entangled). Similarly,  even if each single-particle wavefunction $\phi_a$ and $\phi_b$ is Gaussian, $\psi_N$ will, in general,  be non-Gaussian due to the exchange symmetry  (and the corresponding Wigner function will be non-Gaussian also \cite{PhysRevA.87.022118}). However, there has been much discussion on whether this ``particle'' entanglement is really physical \cite{ECKERT200288,ghirardi2002entanglement,PhysRevA.70.012109,Tichy_2011,EntanglementofIdenticalParticles,2019arXiv190811735M}.

\subsection{Quantum and classical gravity in a single BEC} \label{app:HamBEC}

Using the Newtonian limit of gravity, the QG and CG interaction Hamiltonians for a BEC are given by \eqref{eq:HintQGNewt}-\eqref{eq:HintCGNewt} with $\hat{\Psi}(\bs{r})$ representing the field of the BEC. Taking the limit of zero temperature as in the main text and neglecting any explicit time dependence of the density of the trapped BEC  due to gravity, we can set $\hat{\Psi}(\bs{r}) = \psi(\bs{r}) \hat{a}$, where  $\psi(\bs{r})$ is the condensate wavefunction and $\hat{a}$ is its annihilation operator. This then results in equations \eqref{eq:QGBEC}-\eqref{eq:CGBEC} used in the main text for the interaction Hamiltonians of QG and CG in a single BEC. 

\section{Experimental details of the Bose-Einstein condensate test} \label{app:BEC}

\subsection{Creating the non-classical initial states} 

In Section \ref{sec:Test}, we described how to create an initial single-mode squeezed state of a BEC using a spin-1 BEC and then look for non-Gaussianity. Another option would be to create a single-mode cat state and look for changes in non-Gaussianity. Here, we describe how such a state could be created. Approximating the quantum field of a Bose gas by $\hat{\Psi} = \psi(\bs{r}) \hat{a}$, where $\psi$ is the condensate wavefunction and $\hat{a}$ is the annihilation operator for the condensate, the Hamiltonian for the electromagnetic interactions between the atoms is:
\begin{align} \label{eq:Hcat}
\hat{H} = \hbar \kappa \hat{a}\da \hat{a} \da \hat{a} \hat{a}, 
\end{align} 
where $\kappa := \lambda_{s} / (2 \hbar)$ and $\lambda_{s}$ is defined in \eqref{eq:lambdas}. This Hamiltonian is the Kerr interaction of quantum optics, which has been considered in BECs (see e.g.\ \cite{PhysRevLett.99.010401,PhysRevA.77.023619}). It is known that this Hamiltonian can, in principle, create a Yurke-Stoler state $|\psi \rangle = (|\alpha \rangle + i | - \alpha \rangle)/ \sqrt{2}$ from an initial coherent state $| \alpha \rangle$ \cite{KnightBook}. The evolution of such a state under QG in a BEC is considered in Section \ref{app:k4}.

\subsection{Measuring non-Gaussianity} \label{app:SAC}

As described in Section \ref{sec:Test}, measuring quadrature non-Gaussianity with homodyne or heterodyne detection requires single-atom detection in a quantum gas with high efficiency on small length scales. Recent advances have opened up three promising approaches to this:

\begin{enumerate}
	
	\item After the interaction time $t$, the atomic evolution can be frozen by quickly ramping up a far-detuned optical lattice that confines atoms with a spatial resolution of the lattice wavelength, after which fluorescence-imaging light emitted by the atoms upon exposure to near-resonant light fields can be detected to achieve single atom, high spatial resolution imaging. Single-atom resolved imaging of a quantum gas in a two-dimensional optical lattice with sub-micrometer lattice spacing has been first demonstrated in \cite{bakr2009quantum,Bakr547,sherson2010single}. 

\item  A related optical fluorescence technique follows a similar working principle measuring the transit of single atoms through a light sheet that is located below the atomic sample. While the atoms are falling through the light sheet, a CCD camera records the fluorescence traces. This has been used to measure Hanbury Brown and Twiss correlations across the Bose-Einstein condensation threshold \cite{perrin2012hanbury}.

\item Alternatively, a high finesse cavity can be used where the transit of single atoms through the cavity will cause detectable shifts in the cavity resonance. While this technique does not allow the detection of individual atoms, the emerging photons from the cavity can be used to probe the system, revealing atom number fluctuations in real-time \cite{Brennecke11763,RevModPhys.85.553}. Such techniques have been used to demonstrate many-body entanglement \cite{Haas180,McConnell_2016}.

\end{enumerate}

\section{Fourth-order cumulant for a single-mode bosonic system} \label{app:k4}

The fourth-order cumulant $k_4$ is given by \eqref{eq:k4} for the generalised quadrature $\hat{q} = \hat{a} e^{-i \varphi} + \hat{a}\da e^{i \varphi}$. This requires the determination of various expectation values of combinations of $\hat{a}$ and $\hat{a}\da$:
\begin{align} \label{eq:k4Av}
\braket{\hat{q}^4} = \frac{1}{4} (3+ \braket{\hat{a}^4} e^{-4 i \varphi} + 4 \braket{\hat{a}\da \hat{a}^3} e^{-2 i \varphi} + 6 \langle \hat{a}^2 \rangle e^{-2 i \varphi} + 12 \braket{\hat{a}\da \hat{a}}    + 6 \braket{\hat{a}^{\dagger 2} \hat{a}^2} + h.c.).
\end{align}
The  QG Hamiltonian for a single-mode BEC with electromagnetic inter-atomic interactions neglected is given by \eqref{eq:QGBEC}:
\begin{align} \label{eq:fullHQG}
\hat{H}_{QG} = \hbar \omega \hat{a}\da \hat{a} + \frac{1}{2}  \lambda_{QG} \hat{a}\da \hat{a}\da \hat{a} \hat{a},
\end{align}
where we have also included the free Hamiltonian term $\hbar \omega \hat{a}\da \hat{a}$, which derives from the kinetic  and (time-independent) trapping potential terms of the BEC Hamiltonian (see \eqref{eq:HBECFree}). Working in the Heisenberg picture, the evolution of $\hat{a}$ is:
\begin{align}
\frac{d \hat{a} (t)}{dt} &= - \frac{i}{\hbar} [\hat{a},\hat{H}]\\
&= -i( \omega -  \chi \hat{N}) \hat{a} (t),
\end{align}
where $\hat{N} := \hat{a}\da \hat{a}$  and $\chi := |\lambda_{QG}|/\hbar$. Since $\hat{N}$ is a constant of motion, this can be solved as:
\begin{align}
\hat{a}(t) = e^{-i \omega t} e^{ i \chi \hat{N} t} \hat{a},
\end{align}
where $\hat{a} := \hat{a} (t=0)$. From now on we will ignore the phase  $\omega$ of the free evolution since this can just be absorbed into the $\varphi$ angle of the quadrature $\hat{q}(\varphi) = \hat{a} e^{-i \varphi} + \hat{a}\da e^{i \varphi}$. Then, $\hat{a}^n$  evolves as:
\begin{align} \label{eq:an}
\hat{a}^n(t) &= e^{ i \frac{n}{2} (n-1) \chi t} e^{i n \chi  \hat{N} t} \hat{a}^n,
\end{align}
and therefore:
\begin{align} \label{eq:aman1}
\hat{a}^{\dagger m} \hat{a}^n(t) &= e^{ i \frac{1}{2} (n-m)(m + n - 1) \chi t} \hat{a}^{\dagger m}  e^{i (n-m) \chi \hat{N} t} \hat{a}^n,
\\\label{eq:aman2} &\equiv e^{i \frac{1}{2} (n-m)(m - n + 1) \chi t} e^{i (n-m) \chi \hat{N} t} \hat{a}^{\dagger m} \hat{a}^n.
\end{align}
We could now assume $4 \chi N t \ll 1$, with $N := \braket{\hat{N}}$, and expand the exponentials in \eqref{eq:k4Av}, i.e.\ take:
\begin{align}
e^{i n \chi \hat{N} t} = 1 + i n \chi \hat{N} t + \frac{1}{2!} n^2 \chi^2 \hat{N}^2 t^2 + \cdots
\end{align}
to calculate the expectation value of $\hat{a}^n$ etc. In this case, taking  an initial squeezed coherent state $\ket{\xi,\alpha}$ (which is a general pure Gaussian state), $\kappa_4$ is initially vanishing and remains zero if CG acts (see \eqref{eq:CGBEC}), whereas,  under QG (see \eqref{eq:QGBEC} and \eqref{eq:fullHQG} above), $\kappa_4$ evolves as:
\begin{align}\nonumber
\kappa_4(t) = -3 \chi t \sin \nu  \, \sinh^2 (2 r) \, \eta_1 (r,\nu) +  \frac{3}{8} \chi^2 t^2 \Big[\sinh^2 (2r) &\eta_2(r,\nu)+2 |\alpha|^2 \Big(2\sinh^2(2r) \eta_3(r,\nu) + 2 \sinh (2r) \, \eta_4(r,\nu)  \\ \label{eq:k4SqCoh} &+ 8 \sinh 4r \cos 2 \nu \cos \psi - 5 \sinh 6r \sin 2 \nu \sin \psi\Big)\Big] + \cdots
\end{align}
where:
\begin{align}
\xi &:= r e^{i \vartheta},\\ \label{eq:nu}
\nu &:= 2 \varphi - \vartheta,\\
\eta_1(r,\nu) &:= \sinh 2r - \cos \nu \cosh 2r,\\ \nonumber
\eta_2(r,\nu) &:= 6 \sinh^2(2r) + 8 \cos \nu \sinh 2r (5 \cosh 2r - 2) - \cos 2 \nu (23 \cosh 4r - 16 \cosh 2r +9)\\
\eta_3(r,\nu) &:= 2\sinh 4r [\cos \psi (8 \cos 2 \nu - 3) + 5 \cos \nu] + 3 \cos \psi \cos \nu - \cos 2 \nu,\\
\eta_4(r,\nu) &:= \sinh 6 r(3 - 8 \cos 2 \nu - 5 \cos \nu \cos \psi) - \sin \nu \sin \psi (\cos \nu - 10 \sinh 4r).
\end{align}
 In the limit of a coherent state and $N \gg 1$, we obtain the same scaling found in \cite{PhysRevA.87.033839} at $\chi^4$ with $\varphi = \pi/2$, whereas, in the opposite limit of full squeezing, $\kappa_4$ tends to $ 24 \chi t N^3$ when $\nu = \pi/2$, illustrating that the small value of $\chi$ can be compensated for by a large number of atoms.

If, on the other hand, we had chosen an initial Yurke-Stoler state, $\ket{\psi} := (\ket{\alpha} + i \ket{-\alpha})/\sqrt{2}$, then $\kappa_4$ at time $t$ is:
\begin{align}\nonumber
\kappa_4(t) = &- 8 |\alpha|^4 (\cos^4 \varphi + 3 \sin^4 \varphi\, e^{-8 |\alpha|^2})- 16 \chi t |\alpha|^6 \sin 2 \varphi \Big[\cos^2 \varphi - e^{-4 |\alpha|^2}  [3 + \sin^2 \varphi(2-3  e^{-4 |\alpha|^2})]\Big] + \cdots.
\end{align}
In the limit  $N \gg 1$, the first order term scales as $6 \sqrt{3} \chi t N^3$ at $\varphi = \pi/6$, similar to when the initial state is $\ket{\xi}$ as  above.

\subsection{Non-perturbative approach}
 
 We now pursue a non-perturbative approach to how $\kappa_4$ evolves with time. For the Yurke-Stoler state $\ket{\psi} := (\ket{\alpha} + i \ket{-\alpha})/\sqrt{2}$, we can use:
\begin{align}
\bra{\alpha}  e^{i n \chi  \hat{N} t} \ket{\alpha} \equiv \bra{\alpha}:e^{(\cos [ n \chi t]+ i \sin [n \chi t] -1) \hat{N}}:\ket{\alpha} = e^{(\cos [ n \chi t]+ i \sin [n \chi t] -1) |\alpha|^2},
\end{align}
and \eqref{eq:an} and \eqref{eq:aman1}. For a squeezed coherent state $\ket{\xi, \alpha}$, with $\xi := r e^{i \vartheta}$, we can use \eqref{eq:an} and \eqref{eq:aman2} with:
\begin{align} \label{eq:AvCohSq}
\bra{\alpha,\xi}  e^{i n \chi  \hat{N} t} \ket{\xi, \alpha} \equiv \frac{1}{\sqrt{z}} e^{\frac{1}{2} i n \chi t} G_0 \bra{0}\hat{G}_+ \hat{G}_{2+} \hat{G}_3 \hat{G}_{2-} \hat{G}_-\ket{0},
\end{align}
 where:
 \begin{align}
 G_0 &:= \exp(\beta |\alpha|^2 - \frac{1}{2} \Lambda_{+}  \alpha^{\ast 2} - \frac{1}{2} \Lambda_{-} \alpha^2),\\
 \hat{G}_+ &:= \exp([\beta \alpha - \Lambda_+ \alpha^{\ast} ]\hat{a}),\\
 \hat{G}_- &:= \exp([\beta \alpha^{\ast} - \Lambda_- \alpha]\hat{a}\da),\\
 \hat{G}_{2+} &:= \exp(-\frac{1}{2} \Lambda_+ \alpha^{\ast 2} \hat{a}^{\dagger 2}),\\ 
 \hat{G}_{2-} &:= \exp(-\frac{1}{2} \Lambda_- \alpha^{2} \hat{a}^{2}),\\
 \hat{G}_3 &:= : \exp(\beta \hat{a}\da \hat{a}):,\\
 \beta &:= (1-z)/z,\\
 \Lambda_+ &:= i \sinh (2r) \sin (n \chi t) e^{i \vartheta} /z,\\
   \Lambda_+ &:= i \sinh (2r) \sin (n \chi t) e^{-i \vartheta} /z,\\
    z &:= \cos (n \chi t) - i \cosh (2r) \sin (n \chi t)
 \end{align}
 Here we have used the identities $\exp (\theta[A + B]) \equiv \exp(\theta B) \exp ([e^{\theta} - 1] A) \equiv \exp([1 - e^{-\theta}]A) \exp(\theta B)$ when $[A,B] = A$, as well as \cite{barnett2002methods}:
 \begin{align}
 \begin{aligned} \exp \left(\gamma_{+} \hat{K}_{+}+\gamma_{-} \hat{K}_{-}+\gamma_{3} \hat{K}_{3}\right) &=\exp \left(\Gamma_{+} \hat{K}_{+}\right) \exp \left[\left(\ln \Gamma_{3}\right) \hat{K}_{3}\right] \exp \left(\Gamma_{-} \hat{K}_{-}\right) \end{aligned},
 \end{align}
 with:
 \begin{align}
\Gamma_{3}&=\left(\cosh \beta-\frac{\gamma_{3}}{2 \beta} \sinh \beta\right)^{-2} \\ \Gamma_{ \pm}&=\frac{2 \gamma_{ \pm} \sinh \beta}{2 \beta \cosh \beta-\gamma_{3} \sinh \beta} \\ \beta^{2}&=\frac{1}{4} \gamma_{3}^{2}-\gamma_{+} \gamma_{-},
 \end{align}
 and $[\hat{K}_3,\hat{K}_{\pm}] = \pm \hat{K}_{\pm}$, $[ \hat{K}_+, \hat{K}_-] = -2 \hat{K}_3$. For example, using \eqref{eq:AvCohSq}, $\bra{\xi} \hat{a}^4(t) \ket{\xi}$ under $\hat{H}_{QG}$  can be shown to be:
 \begin{align}
 \bra{\xi} \hat{a}^4(t) \ket{\xi}   = \frac{3 e^{-4 i \chi t + 2 i \vartheta} \sinh^2 (2r) }{2^2 [\cos (4 \chi t) - i \cosh(2r) \sin(4 \chi t)]^{5/2}},
 \end{align}
 where we can use $\sqrt{z} \equiv \sqrt{|z|} (z + |z|)/|z + |z||$ to remove the square root of the complex number.

\subsection{Including the reverse process}

Above we have considered the evolution of $\kappa_4$ under the QG Hamiltonian $\hat{H}_{QG}$ and assuming that the BEC is prepared in either a squeezed coherent state or a Yurke-Stoler state. In the main text, we also considered a measurement protocol where we first prepare the BEC state that is required,  let the BEC evolve under QG, and then measure $\kappa_4$ after we have applied the reverse process to that we used to create the initial BEC state. In the Heisenberg picture, $\hat{a}$ then undergoes the following evolutions:
\begin{enumerate}
	\item $\hat{a} \rightarrow \hat{a}' =  \hat{U}\da_{\Psi} \hat{a} \hat{U}_{\Psi}~~\text{at}~~t = 0$,
	\item $\hat{a} \rightarrow \hat{a}''(t) = e^{i \chi \hat{N} t} \hat{a}'~~\text{for}~~ 0< t  < \tau$,
	\item  $\hat{a} \rightarrow \hat{a}'''(\tau) = \hat{U}\da_{-\Psi} \hat{a}_2(\tau) \hat{U}_{-\Psi}~~\text{at}~~t=\tau$,
	\end{enumerate}
where $\hat{U}_{\Psi}$  refers to the unitary that creates the initial state, and  $\hat{U}_{-\Psi}$ is the reverse process. For example, if a squeezed vacuum state is prepared then $\hat{U}_{\Psi} = \exp(r [e^{i \vartheta} \hat{a}^{\dagger 2} - e^{-i \vartheta} \hat{a}^2]/2)$ and $\hat{U}_{-\Psi} = \exp(-r [e^{i \vartheta} \hat{a}^{\dagger 2} - e^{-i \vartheta} \hat{a}^2]/2)$. In this case, in the limit that $\chi \ll 1$, $\kappa_4$ at the end of the process is:
\begin{align}
\kappa_4(\tau) &= \frac{3}{2} \chi \tau \sin[2\nu] \sinh(2r)^2 + \cdots,
\end{align}
with $\nu$ given by \eqref{eq:nu}. In the limit of large $N$, this scales as $\chi \tau N^2$ in contrast to the $N^3$ scaling for the process considered previously, see \eqref{eq:k4SqCoh}. However, the SNR scaling is the same.

\section{Estimation of the signal-to-noise ratio} \label{app:SNR}

The  SNR for measuring the fourth-order cumulant $\kappa_4$ is given by:
\begin{align}
\text{SNR} = \frac{  |\kappa_4|}{\sqrt{\mathrm{Var} (k_4)}},
\end{align}
where $k_4$ is the fourth $k$-statistic. The variance of $k_4$ is given by \cite{kendall1948advanced}:
\begin{align}\nonumber
\mathrm{Var} (k_4) = \frac{\kappa_8}{\mathcal{M}} + \frac{16 \kappa_2 \kappa_6}{\mathcal{M}-1}+\frac{48 \kappa_3 \kappa_5}{\mathcal{M}-1} + \frac{34\kappa^2_4}{\mathcal{M}-1}+\frac{72 \mathcal{M} \kappa_2^2 \kappa_4}{(\mathcal{M}-1)(\mathcal{M}-2)}+\frac{144 \mathcal{M} \kappa_2 \kappa^2_3}{(\mathcal{M}-1)(\mathcal{M}-2)} + \frac{24 \mathcal{M} (\mathcal{M}+1) \kappa_2^4}{(\mathcal{M}-1)(\mathcal{M}-2)(\mathcal{M}-3)} ,
\end{align}
where $\mathcal{M}$ is the number of independent estimations. In the limit $\mathcal{M} \gg 1$, $\mathrm{Var} (k_4) $ becomes:
\begin{align}\nonumber
\mathrm{Var} (k_4) \approx \frac{1}{\mathcal{M}} \Big[\kappa_8 + 16 \kappa_2 \kappa_6+ 48 \kappa_3 \kappa_5 + 34\kappa^2_4+72 \kappa_2^2 \kappa_4+144  \kappa_2 \kappa^2_3 + 24  \kappa_2^4\Big].
\end{align}
The nth-order cumulant $\kappa_n$, can be found using:
 \begin{align}
 \kappa_n = \mu_n - \sum^{n-1}_{m=1} \binom{n-1}{m-1} \mu_{n-m} \kappa_m,
 \end{align}
where $\mu_n := \braket{\hat{q}^n}$ is the nth moment.

Using results from Appendix \ref{app:k4}, in the limit that $\chi \ll 1$, the SNR for the estimation of $\kappa_4$ for a squeezed vacuum state $\ket{\xi}$ is:
\begin{align}
\text{SNR} = \sqrt{6 \mathcal{M}} t \chi \sinh^2 (2r) \frac{|\sin\nu \, (\sinh 2r - \cos \nu \,\cosh 2r)|}{(  \cosh 2 r - \cos \nu \sinh 2 r  )^2}  + \cdots
\end{align}
This is maximized at the angles:
\begin{align}
\varphi = \frac{1}{2} \Big[ \vartheta  \pm \frac{1}{2} \cos^{-1} y\Big], 
\end{align}
where:
\begin{align}
y := \frac{\sinh^2 2r (\sinh^2 2r - 2) \pm 2 \sqrt{2} \sinh 4r}{(\sinh^2 2r + 2)^2},
\end{align}
which results in the above SNR being approximately $ 4.9 \chi t N^2 \sqrt{\mathcal{M}}$ for $N \gg 1$. When $\chi N^2 t$ is not small, this SNR approximation is not so accurate, and instead the results of the previous section can be used to find a non-perturbative solution to SNR. For example, for the BMV proposal values  $d = 200\unit{\mu m}$,  $t= 2\unit{s}$ and  $M = 10^{-14}\unit{kg}$, we find that the maximum SNR for a spherical  $^{133}\mathrm{Cs}$ BEC is approximately  $0.3 \sqrt{\mathcal{M}}$ (with the value of $d$ being used for the radius $R$). At these values,  $\chi N^2 t = \sqrt{2 /\pi}  \phi \approx 0.5$, where $\phi = 0.6$   is the relative phase expected in the BMV experiment when all  distances between the microspheres other than $d$, the smallest possible distance, are ignored.  Therefore, the SNR is still of order $\chi t N^2 \sqrt{\mathcal{M}}$ in this case. If instead the mass is lowered to $M = 10^{-15}\unit{kg}$ then we can  use the approximation that  $\mathrm{SNR} = 4.9 \chi t N^2 \sqrt{\mathcal{M}}$.

For the protocol where we reverse the squeezing operation before the measurement, the SNR is given by:
\begin{align}
\text{SNR} = \sqrt{\frac{3}{2}} \chi \tau |\sin 2 \nu | \, \sinh^2 (2r) + \cdots, 
\end{align}
in the limit that $\chi \ll 1$.

\section{Evolution under classical gravity} \label{app:CG}

Here we consider how a  single BEC evolves under CG compared to  QG. We start with the general Newtonian expressions \eqref{eq:HintQGNewt} and \eqref{eq:HintCGNewt}. Working in the Schr\"{o}dinger picture, for  QG the evolution of our state vector $\ket{\Psi}$ is given by:
\begin{align} \label{eq:QGStateEv}
i \hbar \frac{d \ket{\varPsi(t)}}{dt} &= \hat{H}^{BEC}_{QG} \ket{\varPsi(t)},
\end{align}
where:
\begin{align}
\hat{H}^{BEC}_{QG} &:= \int d^3 \bs{r} \Big[  - \frac{\hbar^2}{2m} \hat{\Psi}\da (\bs{r}) \nabla^2 \hat{\Psi}(\bs{r}) + V(\bs{r}) \hat{\Psi}\da (\bs{r}) \hat{\Psi} (\bs{r}) +
\frac{1}{2} m  : \hat{\Psi}\da (\bs{r}) \hat{\Psi}  (\bs{r})  \hat{\Phi} (\bs{r}): \Big] \\
& = \int d^3 \bs{r} \Big[  - \frac{\hbar^2}{2m} \hat{\Psi}\da (\bs{r}) \nabla^2 \hat{\Psi}(\bs{r}) + V(\bs{r}) \hat{\Psi}\da (\bs{r}) \hat{\Psi} (\bs{r}) -\frac{1}{2} G m^2  \int d^3 \bs{r'}  \frac{ \hat{\Psi}\da (\bs{r}) \hat{\Psi}\da  (\bs{r'}) \hat{\Psi}  (\bs{r})  \hat{\Psi}  (\bs{r'})}{|\bs{r} - \bs{r'}|} \Big],
\end{align}
with $V(\bs{r})$ the trapping potential. In contrast,  for CG, we have:
\begin{align} \label{eq:CGStateEv}
i \hbar \frac{d \ket{\varPsi(t)}}{dt} = \hat{H}_{CG}^{BEC}[\varPsi](t) \ket{\varPsi(t)},
\end{align} 
where:
\begin{align}
\hat{H}_{CG}^{BEC}[\varPsi](t) := \int d^3 \bs{r} \Big[  - \frac{\hbar^2}{2m} \hat{\Psi}\da (\bs{r}) \nabla^2 \hat{\Psi}(\bs{r}) + V(\bs{r}) \hat{\Psi}\da (\bs{r}) \hat{\Psi} (\bs{r}) +
m   \hat{\Psi}\da (\bs{r}) \hat{\Psi}  (\bs{r})  \Phi[\varPsi(t)] (\bs{r}) \Big].
\end{align}
In the Schr\"{o}dinger-Newton  example of CG, this is:
\begin{align}
\hat{H}_{CG}^{BEC}[\varPsi](t)  = \int d^3 \bs{r} \Big[  &- \frac{\hbar^2}{2m} \hat{\Psi}\da (\bs{r}) \nabla^2 \hat{\Psi}(\bs{r}) + V(\bs{r}) \hat{\Psi}\da (\bs{r}) \hat{\Psi} (\bs{r})\\ &-G m^2 \int d^3 \bs{r'}  \frac{ \hat{\Psi}\da (\bs{r}) \hat{\Psi}  (\bs{r})  \bra{\varPsi(t)} \hat{\Psi}\da  (\bs{r'})  \hat{\Psi}  (\bs{r'}) \ket{\varPsi(t)}}{|\bs{r} - \bs{r'}|} \Big].
\end{align}
Note that the evolution of $\ket{\varPsi}$ in CG  is, in general, `non-linear' in that $\ket{\varPsi}$ is needed to determine $\Phi$. This is often referred to as a wavefunction `self-interaction' since, in the first quantization picture, the wavefunction of a single-particle will now interact with itself, something that can never occur in a quantum theory of gravity, where \eqref{eq:QGStateEv} is  said to be `linear'. 

Neglecting any explicit time dependence,  the evolution of $\ket{\varPsi}$ in QG can, in principle, be solved as:
\begin{align}
\ket{\varPsi(t)} = e^{- i \hat{H}^{BEC}_{QG} t / \hbar} \ket{\varPsi(0)}. 
\end{align}
In contrast, it may not be possible to find an analytic solution in CG due to the potential non-linearities.   However, the evolution will still take the form:
\begin{align}
\ket{\varPsi(t)} =  \hat{T} \left\{ e^{- \frac{i}{\hbar} \int^t_0 d\tau\hat{H}_{CG}^{BEC}[\varPsi] (\tau)} \right\} \ket{\varPsi(0)} , 
\end{align}
where $\hat{T}$ is the time-ordering operator.  Despite the  potential non-linearity, since $\hat{H}_{CG}^{BEC}$ is quadratic in matter field operators, it is still a Gaussian process. For example, consider  the  single-mode BEC experiment introduced in the main text where we assume  $\hat{\Psi} (\bs{r}) = \psi(\bs{r}) \hat{a}$. Neglecting the trapping potential and free dynamics, we then have:
\begin{align}\label{eq:PsiCGa}
\ket{\varPsi(t)}= \hat{T} \left \{e^{- \frac{i}{\hbar} \int^t_0 d \tau \lambda_{CG}[\Psi](t) \hat{a}\da \hat{a} } \right\} \ket{\varPsi(0)} ,
\end{align}
with:
\begin{align} \label{eq:lambdaCGt}
\lambda_{CG}[\varPsi](t) =  m \int d^3 \bs{r} |\psi(\bs{r})|^2 \Phi[\varPsi](t,\bs{r}).
\end{align}
Equation \eqref{eq:PsiCGa} can be written as \cite{Schumaker1986317}:
\begin{align}\label{eq:PsiCGa2}
\ket{\varPsi(t)} = e^{- \frac{i}{\hbar} \Lambda_{CG}[\Psi](t) \hat{a}\da \hat{a} } \ket{\varPsi(0)} ,
\end{align} 
where:
\begin{align}
\Lambda_{CG} [\varPsi](t)  := \int^t_0 d \tau \lambda_{CG} [\varPsi] (\tau).
\end{align}
The evolution of $|\varPsi\rangle$ in this case is then, in general, a non-linear Gaussian process. However, it need not always be non-linear. For instance, in the Schr\"{o}dinger-Newton case we have:  
\begin{align} \label{eq:lambdaCGt2}
\lambda_{CG}[\varPsi](t) =  -  G m^2 \bra{\varPsi(t)} \hat{N} \ket{\varPsi(t)}   \int d^{3} \bs{r} d^{3} \bs{r'} \frac{|\psi(\bs{r'})|^2 |\psi(\bs{r})|^2 }{\left|\bs{r}-\bs{r'}\right|},
\end{align}
where $\hat{N} := \hat{a}\da \hat{a}$. Since $\hat{N}$ is a constant of motion (it commutes with $\hat{H}_{CG}^{BEC}$), we   have:
\begin{align} \label{eq:lambdaCGt3}
\lambda_{CG} =  -  G m^2 N   \int d^{3} \bs{r} d^{3} \bs{r'} \frac{|\psi(\bs{r'})|^2 |\psi(\bs{r})|^2 }{\left|\bs{r}-\bs{r'}\right|},
\end{align}
 where $N := \langle \hat{N} \rangle$. Therefore, $| \varPsi(t)\rangle$ evolves as:
\begin{align}\label{eq:PsiCGa3}
\ket{\varPsi(t)} = e^{- \frac{i}{\hbar} \gamma_{CG} \hat{a}\da \hat{a} t } \ket{\varPsi(0)} ,
\end{align}
where:
\begin{align}
\gamma_{CG} := \int d^3 \bs{r} \Big[  - \frac{\hbar^2}{2m} \psi^{\ast}(\bs{r}) \nabla^2 \psi(\bs{r}) + V(\bs{r}) |\psi(\bs{r})|^2 \Big] - \lambda_{CG} ,
\end{align} 
such that $\ket{\varPsi(t)}$  evolves under a  Gaussian phase-shift channel. For example, if the BEC were initially in a coherent state $|\alpha \rangle$, it would then stay a coherent state but with just a time-dependent phase:
\begin{align}
\ket{\varPsi(t)} &=  \ket{\alpha e^{-i \gamma_{CG} t /\hbar}} ,
\end{align}
with $N = |\alpha|^2$.

\section{Stochastic and complex interactions} \label{app:StochInts}

Here we consider matter interacting with a complex, stochastic non-quantum field (non-operator-valued distribution), and why this interaction cannot, in the absence of all other interactions,  turn a Gaussian state into a non-Gaussian state, where the latter is defined as any state that does not belong to the Gaussian convex hull \cite{ConvexHull}.

In the main text, we considered interacting matter with a classical entity $\mathcal{G}$ (a quantity that takes on real and well-defined values) and how this can be distinguished from the quantum version of the interaction.  Taking, for simplicity, matter to be described by a real scalar quantum field $\hat{\phi}$ then, as long as we do not allow  the classical interaction to induce quantum self-interactions of matter, $\mathcal{G}$ and $\hat{\phi}$ can only interact through  Hamiltonian terms that are linear or quadratic in $\hat{\phi}$. That is, the Hamiltonian density of the interaction must be of the form:
\begin{align} \label{eq:Hdensity}
\hat{\mathcal{H}} = s[\hat{\phi}] f[\mathcal{G}] + t[\hat{\phi}] h[\mathcal{G}],
\end{align} 
where $s$ and $t$ are respectively linear and quadratic real functionals of  $\hat{\phi}$; and $f$ and $h$ are general real functionals of $\mathcal{G}$. It was shown in the main text that a Hamiltonian density of the form \eqref{eq:Hdensity} preserves the Gaussianity of the matter field, and we can use this fact to distinguish it from a quantum interaction.

We now, in contrast to the main text,   allow $\mathcal{G}$, or $f$ and $h$, to be complex-valued. Expanding $\hat{\phi}$ in creation and annihilation operators, $\hat{\phi} = \sum_k [u_k (t) \hat{a}_k + v_k(t) \hat{a}_k\da]$, the corresponding Hamiltonian will be of the form of \eqref{eq:Hquad} except that now $\bs{\lambda}_k(t),  \bs{\mu}_{kl} (t) \in \mathbb{C}$ so that the Hamiltonian is, in general, non-Hermitian. Despite this, the quadratic nature of the non-Hermitian Hamiltonian means that it still preserves the Gaussian form of the state for an initial Gaussian state \cite{PhysRevA.83.060101,Graefe_2012,Graefe_2015}\cite{Note16}. For example, consider the Hamiltonian $\hat{H} = \lambda \hat{a}\da \hat{a}$, where $\lambda := \lambda_R - i \lambda_I$. Under this Hamiltonian,  an initial coherent state $|\alpha\rangle$ will evolve to $\exp\{-|\alpha|^2 (1 - \exp\{-2 \lambda_I t\})/2\} |\alpha\exp\{-i\lambda t\}\rangle$, which is just an unnormalized, damped coherent state with a time-dependent phase (note that we have taken $\hbar = 1$ here and do so throughout the rest of this Appendix). In fact, in general,  a non-Hermitian Hamiltonian will lead to an unnormalized state. To rectify this, the physical state vector can be defined as $|\psi^{\mathcal{N}}\rangle := |\psi\rangle / ||\psi\rangle|$. For the above example, this would mean that an initial coherent state  evolves to a damped coherent state with a time-dependent phase: $|\alpha'(t) \exp\{-\lambda_I t\}\rangle$, where $\alpha'(t) := \alpha \exp\{-i \lambda_R t\}$.

We now take $\mathcal{G}$ to be a stochastic field, which we denote as $\tilde{\mathcal{G}}$, and keep $f$ and $h$  complex-valued. The interaction Hamiltonian density \eqref{eq:Hdensity} can then be written as:
\begin{align} \label{eq:StochclassicalG}
\hat{\mathcal{H}}[\tilde{\mathcal{G}}] = s[\hat{\phi}] f[\tilde{\mathcal{G}}] + t[\hat{\phi}] h[\tilde{\mathcal{G}}].
\end{align}
 In the interaction picture, an out state $|\psi_{out} [\tilde{\mathcal{G}}] \rangle$ of the quantum field $\hat{\phi}$  is now  given by a stochastic S-matrix $\hat{S}[\tilde{\mathcal{G}}]$ acting on the in state $|\psi_{in}  \rangle$ \cite{PhysRevA.42.5086}. That is:
\begin{align} \label{eq:psiout}
|\psi_{out} [\tilde{\mathcal{G}}] \rangle = \hat{S} [\tilde{\mathcal{G}}]|\psi_{in}  \rangle,
\end{align}
where:
\begin{align}
\hat{S}[\tilde{\mathcal{G}}] := T e^{-i \int d^4 x (\hat{\mathcal{H}}_0 + \hat{\mathcal{H}}[\tilde{\mathcal{G}}])}, 
\end{align}
with $T$ the time-ordering operator;  $x$  a four-coordinate; and $\hat{\mathcal{H}}_0$ the free (non-stochastic) Hamiltonian density.

Since the Hamiltonian may not be Hermitian, the out state may not be normalized, but we can define a normalized out state as:
\begin{align}
|\psi^{\mathcal{N}}_{out} [\tilde{\mathcal{G}}] \rangle := \mathcal{N}^{-1/2} |\psi_{out} [\tilde{\mathcal{G}}] \rangle
\end{align} 
where:
\begin{align}
\mathcal{N} := \langle \psi_{out} [\tilde{\mathcal{G}}] |\psi_{out} [\tilde{\mathcal{G}}] \rangle.
\end{align}
The density matrix corresponding to a particular out state $|\psi_{out} [\tilde{\mathcal{G}}] \rangle$ can be defined as usual:
\begin{align}
\hat{\rho}_{out} [\tilde{\mathcal{G}}] :=|\psi_{out} [\tilde{\mathcal{G}}] \rangle \langle \psi_{out} [\tilde{\mathcal{G}}] |,
\end{align}
or the normalized version:
  \begin{align} \label{eq:normrho}
  \hat{\rho}^{\mathcal{N}}_{out} [\tilde{\mathcal{G}}] := \mathcal{N}^{-1} \hat{\rho}_{out} [\tilde{\mathcal{G}}] .
  \end{align}
From  \eqref{eq:psiout}, the density matrix $\hat{\rho}_{out} [\tilde{\mathcal{G}}]:=|\psi_{out} [\tilde{\mathcal{G}}]\rangle \langle\psi_{out} [\tilde{\mathcal{G}}]|$ can be found through:
\begin{align}
\hat{\rho}_{out} [\tilde{\mathcal{G}}] &=\hat{S} [\tilde{\mathcal{G}}] \hat{\rho}_{in} \hat{S}\da [\tilde{\mathcal{G}}],
\end{align} 
where $\hat{\rho}_{in} := |\psi_{in}\rangle \langle \psi_{in} |$. The above density matrix corresponds to a particular stochastic out state $|\psi_{out}[\tilde{\mathcal{G}}]\rangle$. However, the quantity that provides the correct expectation values of operators  ($\langle \hat{A} \rangle = \mathrm{Tr} [\hat{\rho}_{out} \hat{A}]$)  is the average density matrix (averaged over $\tilde{\mathcal{G}}$) $\hat{\rho}_{out}$ \cite{OpenBook,DIOSI1985288}. That is, $\hat{\rho}_{out}$ is given by \cite{PhysRevA.42.5086}:
\begin{align} \label{eq:rhoout}
\hat{\rho}_{out} &:= \int \mathcal{D} \tilde{\mathcal{G}} \, P [\tilde{\mathcal{G}}]\, (\hat{\rho}_{out} [\tilde{\mathcal{G}}])\\
&= \int \mathcal{D} \tilde{\mathcal{G}}\, P [\tilde{\mathcal{G}}]\, (\hat{S}[\tilde{\mathcal{G}}] \, \hat{\rho}_{in} \, \hat{S}\da[\tilde{\mathcal{G}}])\\
&:=\int \mathcal{D} \tilde{\mathcal{G}}\, P [\tilde{\mathcal{G}}]\, (\hat{\mathcal{S}}_S[\tilde{\mathcal{G}}] \, \hat{\rho}_{in})\\
&:= \hat{\mathcal{S}}_{av} \hat{\rho}_{in}
\end{align}
where $P[\tilde{\mathcal{G}}]$ is the probability distribution functional of $\tilde{\mathcal{G}}$; $\hat{\mathcal{S}}_S[\tilde{\mathcal{G}}]$ is the scattering superoperator; and $\hat{\rho}_{in}$ now, in general, corresponds to a general initial mixed state \cite{PhysRevLett.52.1657,Gisin1989}.

Taking $\hat{\rho}_{in}$ to be a pure Gaussian state, then since each $\hat{S}[\tilde{\mathcal{G}}]$ is associated with a Gaussian  transformation (i.e.\ \eqref{eq:Hdensity}), \eqref{eq:rhoout} is just the stochastic quantum field theory generalization of a state, $\hat{\rho}_{ch}$, in the Gaussian convex hull of quantum optics \cite{ConvexHull}:
\begin{align} \label{eq:GConvexHull}
\hat{\rho}_{ch} = \int d \bs{g} \, P(\bs{g})\, \hat{\rho}_G (\bs{g}),
\end{align}
where $\bs{g}$ is a set of complex numbers, $P(\bs{g})$ is a probability distribution, and $\hat{\rho}_G (\bs{g}) = |\psi_G (\bs{g}) \rangle \langle \psi_G (\bs{g}) |$ is a pure Gaussian density matrix. Defining a pure Gaussian state as a pure state fully defined by its first and second moments, a  mixture of such states  represents a broader definition of a Gaussian state compared to  the more conventional definition of any state that is fully characterized by its first and second moments, which is  used in the main text \cite{ConvexHull,PhysRevA.90.013810}. A non-Gaussian state (also sometimes referred to as a `quantum' non-Gaussian state to distinguish it from the more conventional definition of a non-Gaussian state \cite{PhysRevA.90.013810}) can then be defined as any state that lives outside the convex hull of Gaussian states\footnotetext[26]{Note that a  state in the Gaussian convex hull  can have an overall non-Gaussian Wigner function.} \cite{Note26}.

As shown and discussed in the main text, to rule out a classical interaction (defined as an interaction with a non-quantum field that takes on real and well-defined values, such as the classical electromagnetic or gravitational fields) any detection of a non-Gaussian state as it is conventionally defined (any state not fully defined by its first and second moments),  is sufficient as long as all other interactions can be neglected. As shown above, this also applies when the field takes on complex values. However, to rule out a stochastic interaction (defined as an interaction with a non-quantum field that is fundamentally stochastic, sometimes referred to as a `post-quantum' interaction), we must appeal to the detection of a non-Gaussian state (or `quantum' non-Gaussian state) in its broader definition as any state that sits outside the Gaussian convex hull\footnotetext[27]{Here we are assuming, as in the main text, that the interaction is of the form \eqref{eq:Hdensity} or, when extended to stochastic fields, \eqref{eq:StochclassicalG}.} \cite{Note27}. This is to be expected since a Gaussian state evolves to a state in the Gaussian convex hull if there is a combination of Gaussian operations and statistical randomization \cite{PhysRevA.87.062104}.

\subsection{Example: A  stochastic and complex generalization of the Newtonian gravitational interaction}

We now  consider a specific example of a stochastic and complex interaction that, when we take the non-relativistic limit,  could be considered   as a stochastic and complex generalization of the Newtonian gravitational interaction. The relativistic version of this interaction has the   Hamiltonian density:
\begin{align} \label{eq:EgHamDens}
\hat{\mathcal{H}} =
\hat{A}\, \tilde{h}[\tilde{\mathcal{G}}],
\end{align}
where $\hat{A} := \hat{\phi}\da \hat{\phi}$ is a mass-density-like operator for a complex relativistic scalar field $\hat{\phi}$, and $\tilde{h}[\tilde{\mathcal{G}}]$ is defined as:
\begin{align}
\tilde{h}[\tilde{\mathcal{G}}(x)] :=  \int d^4 x' \Lambda (x,x') \tilde{\mathcal{G}} (x'),
\end{align}
with $\Lambda(x,x') := \Lambda_R(x,x') - i \Lambda_I(x,x')$; $\Lambda_{R}(x,x')$ a real kernel; $\Lambda_{I}(x,x')$ a positive definite  kernel; and $\tilde{\mathcal{G}}(x)$ a real stochastic field. The stochastic (Gaussian) scattering matrix $\hat{S}[\tilde{\mathcal{G}}] $ is then (ignoring $\hat{\mathcal{H}}_0$ for simplicity):
\begin{align}
\hat{S}[\tilde{\mathcal{G}}] = T e^{-i \int d^4 x d^4 x' \Lambda_R(x,x') \hat{\phi}\da(x') \hat{\phi}(x') \tilde{\mathcal{G}}(x')} e^{-\int d^4 x d^4 x' \Lambda_I(x,x') \hat{\phi}\da(x') \hat{\phi}(x') \tilde{\mathcal{G}}(x')},
\end{align} 
such that the (Gaussian) stochastic scattering superoperator is:
\begin{align}
\hat{\mathcal{S}}_{S}[\tilde{\mathcal{G}}] = \hat{T} \exp\left\{-i \int d^4 x d^4 x' \Lambda_R(x,x') \hat{A}_{\Delta}(x') \tilde{\mathcal{G}}(x') -\int d^4 x d^4 x' \Lambda_I(x,x') \hat{A}_{\Sigma}(x') \tilde{\mathcal{G}}(x')\right\},
\end{align} 
where   $\hat{T}$ is the time-ordering superoperator;  $\hat{A}_{\Delta} = \hat{A}_+ - \hat{A}_-$, and $\hat{A}_{\Sigma} = \hat{A}_+ + \hat{A}_-$, with $\hat{A}_+$ representing $\hat{A}$ acting on $\hat{\rho}_{in}$ from the left, and $\hat{A}_-$ representing $\hat{A}$ from the right. Taking, for convenience, the probability distribution functional to be Gaussian:
\begin{align}
P[\tilde{\mathcal{G}}] = (\det \Gamma)^{1/2} \, e^{-\int d^4 x d^4 x' \Gamma(x,x')  \tilde{\mathcal{G}}(x) \tilde{\mathcal{G}}(x')},
\end{align}
with $\Gamma(x,x')$ a positive-definite symmetric kernel, then we can perform  Gaussian functional integration \eqref{eq:rhoout} over $\tilde{\mathcal{G}}$ to obtain $\hat{\rho}_{out} = \hat{\mathcal{S}}_{av} \hat{\rho}_{in}$, with $\hat{\mathcal{S}}_{av}$:
\begin{align}
\hat{\mathcal{S}}_{av} = \hat{T} e^{\int d^4x d^4x' [-\beta_{RR}(x,x') \hat{A}_{\Delta}(x) \hat{A}_{\Delta}(x') + i \beta_{IR}(x,x') \hat{A}_{\Delta}(x) \hat{A}_{\Sigma}(x') + i\beta_{IR}(x,x') \hat{A}_{\Sigma}(x) \hat{A}_{\Delta}(x') + \beta_{II}(x,x') \hat{A}_{\Sigma}(x) \hat{A}_{\Sigma}(x') ]},
\end{align}
where:
\begin{align}
\beta_{RR}(x,x')  &:= \frac{1}{4} \int d^4x'' d^4x''' \Lambda_R(x,x'') \Gamma^{-1} (x'',x''') \Lambda_R(x',x'''),\\ 
\beta_{IR}(x,x') &:= \frac{1}{4} \int d^4x'' d^4x''' \Lambda_R(x,x'') \Gamma^{-1} (x'',x''') \Lambda_I(x',x'''),\\  \label{eq:betaII}
\beta_{II}(x,x') &:= \frac{1}{4} \int d^4x'' d^4x''' \Lambda_I(x,x'') \Gamma^{-1} (x'',x''') \Lambda_I(x',x''').
\end{align}
We now take a Markovian approximation and define $\Lambda_R$,  $\Lambda_I$ and $\Gamma$ as \cite{PhysRevA.42.5086}:
\begin{align}
\Lambda_{R,I}(x,x') &= \lambda_{R,I}(x_0,\bs{r},\bs{r'}) \delta(x_0-x'_0),\\
\Gamma(x,x') &= \gamma(x_0,\bs{r},\bs{r'}) \delta(x_0-x'_0).
\end{align}
The superoperator  $\hat{S}_{av}$ can then be written as:
\begin{align}
\hat{S}_{av} = \hat{T} \exp\left\{ \int^{\infty}_{-\infty} dt \hat{\mathcal{L}} (t) \right\},
\end{align}
where $\hat{\mathcal{L}}(t)$ is the linear evolution superoperator and $\hat{T}$ the time-ordering superoperator \cite{PhysRevA.42.5086}. The averaged density matrix $\hat{\rho}$ at a time $t$ can now be obtained through:
\begin{align}
\hat{\rho}(t) = \hat{T} \exp\left\{ \int^t_{0} d \tau  \hat{\mathcal{L}}(\tau)\right\} \hat{\rho}(0).
\end{align}
  The superoperator $\hat{\mathcal{L}}$ acts on $\hat{\rho}$ as:
\begin{align}\nonumber
\hat{\mathcal{L}} \hat{\rho} = \int d \bs{r} d\bs{r'} \Big[ &2 i b_{IR}(t,\bs{r},\bs{r'}) [\hat{A}(t,\bs{r}) \hat{A}(t,\bs{r'}),\hat{\rho}] - b_{RR}(t,\bs{r},\bs{r'}) [\hat{A}(t,\bs{r}),[\hat{A}(t,\bs{r'}),\hat{\rho}]] \\  &+ b_{II}(t,\bs{r},\bs{r'}) \{  \hat{A}(t,\bs{r}),\{\hat{A}(t,\bs{r'}),\hat{\rho}\}\}\Big]
\end{align} 
where:
\begin{align}
b_{RR}(t,\bs{r},\bs{r'})  &:= \frac{1}{4} \int d \bs{r''} \bs{r'''} \lambda_R(t,\bs{r},\bs{r''}) \gamma^{-1} (t,\bs{r''},\bs{r'''}) \lambda_R (t,\bs{r'},\bs{r'''}),\\ 
b_{IR}(t,\bs{r},\bs{r'})  &:= \frac{1}{4} \int d \bs{r''} \bs{r'''} \lambda_I(t,\bs{r},\bs{r''}) \gamma^{-1} (t,\bs{r''},\bs{r'''}) \lambda_R (t,\bs{r'},\bs{r'''}),\\
b_{II}(t,\bs{r},\bs{r'})  &:= \frac{1}{4} \int d \bs{r''} \bs{r'''} \lambda_I(t,\bs{r},\bs{r''}) \gamma^{-1} (t,\bs{r''},\bs{r'''}) \lambda_I (t,\bs{r'},\bs{r'''}),
\end{align}
and we have used $[\hat{A}^2,\hat{\rho}] \equiv  \{\hat{A},[\hat{A},\hat{\rho}]\} \equiv [\hat{A},\{\hat{A},\hat{\rho}\}]$. Therefore, $\hat{\rho}(t)$ obeys the following master equation:
\begin{align} \nonumber
\frac{d \hat{\rho}(t)}{dt} =  \int d \bs{r} d\bs{r'} \Big[ &2i b_{IR}(t,\bs{r},\bs{r'}) [\hat{A}(t,\bs{r}) \hat{A}(t,\bs{r'}),\hat{\rho}] - b_{RR}(t,\bs{r},\bs{r'}) [\hat{A}(t,\bs{r}),[\hat{A}(t,\bs{r'}),\hat{\rho}]] \\\label{eq:RelMasterEq} &+ b_{II}(t,\bs{r},\bs{r'}) \{  \hat{A}(t,\bs{r}),\{\hat{A}(t,\bs{r'}),\hat{\rho}\}\}\Big].
\end{align}
Finally, we take the non-relativistic limit and  replace $\hat{\phi}$ with the non-relativistic scalar field $\hat{\Psi}$. The Hamiltonian density can then be thought of as in the form of that for a stochastic and complex generalization  of the  Newtonian gravitational interaction, with $h[\tilde{G}]$ a complex and stochastic generalization of the Newtonian potential, and matter  represented by $\hat{\Psi}$.  Assuming  that $\beta_{RR}$, $\beta_{IR}$ and $\beta_{RR}$ are time-independent, we  end up with:
\begin{align} \nonumber
\frac{d \hat{\rho}(t)}{dt} = \int d \bs{r} d\bs{r'} \Big( &2i b_{IR}(\bs{r},\bs{r'}) [\hat{\Psi}\da(\bs{r}) \hat{\Psi}(\bs{r}) \hat{\Psi}\da(\bs{r'}) \hat{\Psi}(\bs{r'}),\hat{\rho}(t)] - b_{RR}(\bs{r},\bs{r'}) [\hat{\Psi}\da(\bs{r}) \hat{\Psi}(\bs{r}),[\hat{\Psi}\da(\bs{r'}) \hat{\Psi}(\bs{r'}),\hat{\rho}(t)]] \\ \label{eq:fullmaster} &+ b_{II}(\bs{r},\bs{r'}) \{  \hat{\Psi}\da(\bs{r}) \hat{\Psi}(\bs{r}),\{\hat{\Psi}\da(\bs{r'}) \hat{\Psi}(\bs{r'}),\hat{\rho}(t)\}\}\Big),
\end{align}
where we have ignored the time dependence of $\hat{\Psi}$ for simplicity. The first term is of the same form as that  which would be induced by the Newtonian limit of QG (see Appendix \ref{app:NewtLimitQG}). However, despite Newtonian QG inducing non-Gaussianity, the other two terms conspire with the first  to reduce the full process to  a channel that keeps a Gaussian state in the (unnormalized) Gaussian convex hull. That is, despite the appearance of the first term, this master equation cannot turn a Gaussian state into a non-Gaussian state (defined as a state that lives outside the Gaussian convex hull). This is clear from our starting point \eqref{eq:rhoout} for the averaged density matrix $\hat{\rho}$. 

We can, in fact, write the solution of \eqref{eq:fullmaster} as a state in the standard quantum optics definition of the Gaussian convex hull by, for example, dropping the temporal and spatial dependence of $\tilde{\mathcal{G}}$,  $\gamma$,  $\lambda_R$ and $\lambda_I$:  the above master equation can then be written as: 
\begin{align} \label{eq:simplermasterEq}
\frac{d \hat{\rho}(t)}{dt} = -i \kappa_{IR}  [(\hat{a}\da \hat{a})^2 ,\hat{\rho}(t)] - \kappa_{RR}[\hat{a}\da \hat{a},[\hat{a}\da \hat{a},\hat{\rho}(t)]] + \kappa_{II} \{  \hat{a}\da \hat{a},\{\hat{a}\da \hat{a} ,\hat{\rho}(t)\}\},
\end{align}
where we have also taken the single-mode approximation $\hat{\Psi} (\bs{r}) = \psi(\bs{r}) \hat{a}$ found in the main text, and defined $
\kappa_{RR} := \frac{1}{4} \kappa^2 \lambda_R^2$; $\kappa_{IR} :=  \frac{1}{2} \kappa^2 \lambda_I \lambda_R$; $\kappa_{II} := \frac{1}{4} \kappa^2 \lambda_I^2$; $\kappa := \int d \bs{r} |\psi(\bs{r})|^2$ and  $\gamma = \delta^{(3)}(\bs{r}-\bs{r'})$ for convenience. Using \eqref{eq:rhoout}, the solution to \eqref{eq:simplermasterEq} can be written as:
\begin{align} \label{eq:rhotconvex}
\hat{\rho}(t) =  \int d g P(g,t) e^{-i \kappa \lambda_{R} \hat{a}\da\hat{a} g t - \kappa \lambda_{I} \hat{a}\da \hat{a} g t } \hat{\rho}(0) e^{i \kappa \lambda_{R} \hat{a}\da\hat{a} g t - \kappa \lambda_I \hat{a}\da \hat{a} g t },
\end{align}
with:
\begin{align}
P(g,t) := \sqrt{\frac{t}{ \pi}} e^{- g^2 t},
\end{align}
where $g \in \mathbb{R}$ is a dummy variable used in place of $\tilde{\mathcal{G}}$. If $\hat{\rho}(0)$ in \eqref{eq:rhotconvex} is a pure Gaussian state, the density matrix $\hat{\rho}(t)$ of  \eqref{eq:rhotconvex}, which solves \eqref{eq:simplermasterEq},  is then part of the (in general, non-normalized) Gaussian convex hull \eqref{eq:GConvexHull}.

\subsection{Preserving the norm: relationship to objective collapse theories and continuous-time measurements} \label{app:PreservingNorm}

The master equations \eqref{eq:RelMasterEq} and \eqref{eq:fullmaster}  (and so also \eqref{eq:simplermasterEq}) do not preserve the norm of the state. As detailed above, in order to preserve the norm, each stochastic density matrix can be redefined through \eqref{eq:normrho} and we can then take these as the physical  stochastic density matrices. However, this  results in a non-linear evolution of the new averaged  density matrix $\hat{\rho}$, which can lead to superluminal signalling \cite{Gisin1989,PhysRevA.42.5086}. This issue can also be found in objective-collapse theories  where matter is coupled to a stochastic field through an anti-Hermitian term involving a particular matter operator $\hat{A}$ \cite{adler2004quantum,PhysRevD.96.104013}. In these models a term of the form $\hat{A}^2$ is  included in the evolution of the stochastic state vector  in order to eliminate the problematic non-linear terms in the evolution of the averaged density matrix \cite{GRW1,GRW2,Diosi1989,CSL,PhysRevA.42.5086,adler2004quantum,Adler_2007,PhysRevD.96.104013}. Such higher-order terms can also be used to eliminate the non-norm preserving terms in the evolution of the non-normalized density matrix \cite{PhysRevA.42.5086}. For example, to our   Hamiltonian density \eqref{eq:EgHamDens}, we can include a term of the form $\hat{A}^2$:
\begin{align} \label{eq:HSelfInt}
\hat{\mathcal{H}}[\tilde{\mathcal{G}}(x)] &:= \int d^4 x' \Lambda(x,x') \tilde{\mathcal{G}}(x') \hat{A}(x) -  2 i \int d^4 x'  \beta_{II} (x,x')  \hat{A}(x') \hat{A}(x),
\end{align}
with $\hat{A}(x) := \hat{\phi}\da(x) \hat{\phi}(x)$ and $\beta_{II}$ defined in \eqref{eq:betaII}. Taking the Markovian limit and assuming a Gaussian profile for $\tilde{\mathcal{G}}$ as above, the new term turns the non-norm preserving term in \eqref{eq:RelMasterEq} into a norm-preserving term:
\begin{align} 
\frac{d \hat{\rho}(t)}{dt} =  \int d \bs{r} d\bs{r'} \Big( &2i b_{IR} [\hat{A}(t,\bs{r}) \hat{A}(t,\bs{r'}),\hat{\rho}(t)] - (b_{RR}+b_{II})  [\hat{A}(t,\bs{r}),[\hat{A}(t,\bs{r'}),\hat{\rho}(t)]] \Big).
\end{align} 
Since $\hat{A} := \hat{\phi}\da \hat{\phi}$, the new term in \eqref{eq:HSelfInt}  is an (anti-Hermitian) quantum self-interaction of matter. That is, we have effectively introduced a new force. This new quantum force will, in general, induce non-Gaussianity. However, if we take $\Lambda_R = 0$ in \eqref{eq:HSelfInt} so that the stochastic interaction is anti-Hermitian (as is usually the case in objective-collapse theories), then the master equation simplifies to:
\begin{align} \label{eq:GaussMaster}
\frac{d \hat{\rho}(t)}{dt} =   - \int d \bs{r} d\bs{r'} b_{II} (t,\bs{r},\bs{r'}) [\hat{A}(t,\bs{r}),[\hat{A}(t,\bs{r'}),\hat{\rho}(t)]].
\end{align}
which is a master equation that preserves the Gaussian convex hull since such a master equation is also derived when taking $\Lambda_I = 0$ in the original theory without the new quantum self-interaction force (see \eqref{eq:RelMasterEq} with $b_{IR}=b_{II}=0$). When taking the non-relativistic limit $\hat{\phi} \rightarrow \hat{\Psi}$, \eqref{eq:GaussMaster} is of the form of the master equation found in objective-collapse theories such as continuous spontaneous localization (CSL) and Di\'{o}si-Penrose\footnotetext[30]{{Taking $\Lambda_R = 0$, $\Lambda_I(x,x') =: -\sqrt{\lambda} \delta^{(4)}(x-x')$, $\Gamma =\delta^{(4)}(x-x')$ and replacing $\hat{A}$ with $(\hat{A} - \langle \hat{A} \rangle)$, the Hamiltonian density \eqref{eq:HSelfInt} becomes:
\begin{align} 
-i\hat{\mathcal{H}}[\tilde{\mathcal{G}}] = \sqrt{\lambda}  (\hat{A} - \langle \hat{A} \rangle)  \tilde{\mathcal{G}} -  \frac{1}{2} \lambda (\hat{A} - \langle \hat{A} \rangle)^2,
\end{align}	
which is closely connected to the evolution of the state vector found in objective-collapse theories such as CSL and Di\'{o}si-Penrose \cite{CSL,Diosi1989,PhysRevD.96.104013}.}} \cite{Diosi1989,CSL}\cite{Note30}. It is also the master equation of continuous-time measurements in the basis $\hat{A}$, such that we can essentially consider the stochastic field $\tilde{\mathcal{G}}$ and new quantum self-interaction force $\beta_{II} \hat{A}^2$ working together to perform continuous measurements  of matter (that preserve the Gaussian convex hull).

If, however, both $\Lambda_R$ and $\Lambda_I$ are non-zero (see e.g.\ \cite{PhysRevA.42.5086,PhysRevD.96.104013} for similar models), then, in general, the non-Gaussian character of the new quantum self-interaction force $\beta_{II} \hat{A}^2$ is preserved, and we have a channel that can induce non-Gaussianity.    Even so, in the asymptotic limit, the state will become a state of the Gaussian convex hull rather than a non-Gaussian state. 

When both $\Lambda_R$ and $\Lambda_I$ are non-zero (and we also have the new quantum self-interaction force $\beta_{II} \hat{A}^2$), the theory is closely related to  a continuous-time measurement being performed by the two interactions as above but now with a feedback mechanism \cite{PhysRevA.42.5086}. Note that weak measurements with local feedback operations can also induce entanglement in the case of joint  measurements \cite{PhysRevA.77.014305,PhysRevA.92.062321}.


\twocolumngrid

\end{document}